\documentclass[12pt]{article}
\usepackage{amssymb}
\usepackage{graphicx}
\usepackage{amsmath}

\def\eeq{\end{eqnarray}}

\def\D{\mathcal{D}}

\def\=:{=\hspace{-.7em}\raisebox{1.1ex}{.}\hspace{.1em}\raisebox{-0.2ex}{.} }

\newcommand {\beq}{\begin{eqnarray}}
\newcommand {\eeqq}{\end{eqnarray}}

\newcommand {\1}[1]{\frac{1}{#1}}

\newcommand {\lam}{\lambda}

\newcommand {\Tr}{{\rm Tr}\,}

\makeatletter
\@addtoreset{equation}{section}

\renewcommand{\thefootnote}{\fnsymbol{footnote}}
\makeatother

\begin{document}
\thispagestyle{empty}
\begin{flushright}
TIT/HEP--554, IFUP-TH/2006-13, RIKEN-TH-73, UT-Komaba/06-6\\
{\tt hep-th/0607070} \\
July, 2006 \\
\end{flushright}
\vspace{3mm}

\begin{center}
{\Large \bf
Non-Abelian Vortices of Higher Winding Numbers}
\\[12mm]
\vspace{5mm}

\normalsize
 {\large \bf
Minoru~Eto}$^1$\footnote{\it  e-mail address:
meto@hep1.c.u-tokyo.ac.jp
},
  {\large \bf
Kenichi~Konishi}$^{2,3}$\footnote{\it  e-mail address:
konishi@df.unipi.it
},
  {\large \bf
Giacomo~Marmorini}$^{4,3}$\footnote{\it  e-mail address:
g.marmorini@sns.it
},
  {\large \bf
Muneto~Nitta}$^5$\footnote{\it  e-mail address:
nitta@phys-h.keio.ac.jp
},
  {\large \bf
Keisuke~Ohashi}$^6$\footnote{\it  e-mail address:
keisuke@th.phys.titech.ac.jp
},
  {\large \bf
Walter~Vinci}$^{2,3}$\footnote{\it  e-mail address:
walter.vinci@pi.infn.it
},
  {\large \bf
Naoto~Yokoi}$^{7}$\footnote{\it  e-mail address:
n.yokoi@riken.jp
}

\vskip 1.5em

{\small
$^1$ {\it
University of Tokyo,  Inst. of Physics
Komaba 3-8-1, Meguro-ku
Tokyo 153, Japan
}
\\

$^2$ {\it Department of Physics, University of Pisa \\
Largo Pontecorvo, 3,   Ed. C,  56127 Pisa, Italy
}

$^3$ {\it INFN, Sezione di Pisa, \\
Largo Pontecorvo, 3, Ed. C, 56127 Pisa, Italy
}
\\
$^4$ {\it Scuola Normale Superiore, \\
Piazza dei Cavalieri, 7, 56100
 Pisa, Italy
}
\\
$^5$ {\it
Department of Physics, Keio University, Hiyoshi,
Yokohama, Kanagawa 223-8521, JAPAN
}
\\
$^6$ {\it Department of Physics, Tokyo Institute of
Technology \\
Tokyo 152-8551, JAPAN}
\\
$^7$ {\it
Theoretical Physics Laboratory,\\
The Institute of Physical and Chemical Research (RIKEN),\\
2-1 Hirosawa, Wako, Saitama 351-0198, JAPAN
}

}
%
%

%
{\bf Abstract}\\[5mm]
{\parbox{13cm}{\hspace{5mm}

We make a detailed study of the moduli space of  winding number two
($k=2$)   axially symmetric vortices (or equivalently, of co-axial
composite of two fundamental vortices),   occurring in $U(2)$ gauge
theory with two flavors in the Higgs phase, recently discussed by
Hashimoto-Tong and Auzzi-Shifman-Yung. We find that it is a weighted
projective space $W{\bf C}P^2_{(2,1,1)} \simeq   {\bf C}P^{2}/{\bf Z}_2$. 
This manifold
contains an $A_1$-type (${\bf Z}_2$) orbifold singularity even
though the full moduli space including the relative position moduli
is smooth. The $SU(2)$ transformation properties of such vortices are
studied. Our results are  then generalized to $U(N)$ gauge theory
with $N$ flavors, where  the internal  moduli space of $k=2$
axially symmetric  vortices is found  to be a weighted Grassmannian
manifold. 
It contains singularities along a submanifold.

}}
\end{center}
\vfill
\newpage
\setcounter{page}{1}
\setcounter{footnote}{0}
\renewcommand{\thefootnote}{\arabic{footnote}}

\section{Introduction}\label{INTRO}

Vortices have played important roles in various areas of fundamental
physics since their discovery \cite{Abrikosov:1956sx,Nielsen:cs}. A
particularly interesting type of  vortices are the ones possessing
an exact, continuous   non-Abelian flux moduli (called non-Abelian
vortices below), found recently \cite{Hanany:2003hp,Auzzi:2003fs}. A
motivation for studying such non-Abelian vortices is  that a
monopole is confined in the Higgs phase by related  vortices
\cite{Tong:2003pz}-\cite{Eto:2004rz}, so that such systems provide a
dual model  of color confinement of truly non-Abelian kind. Another
motivation could come from the interest in the physics of cosmic
strings. Many papers on non-Abelian vortices appeared lately
\cite{Isozumi:2004vg}--\cite{Eto:2006pg}, where the discussion often
encompasses the context of more general soliton physics, involving
domain walls, monopoles and vortices, or composite  thereof.

The moduli space of non-Abelian vortices was obtained in certain
D-brane configuration in string theory \cite{Hanany:2003hp} as well
as in a  field theory framework \cite{Eto:2005yh}. In particular,
the moduli subspace of $k=2$  axially symmetric vortices was studied
by two groups: Hashimoto and Tong (HT) \cite{Hashimoto:2005hi} and
Auzzi, Shifman and Yung (ASY) \cite{Auzzi:2005gr}. The former
concluded that it is ${\bf C}P^2$ by using the brane construction
\cite{Hanany:2003hp} whereas the latter  found  ${\bf C}P^2/{\bf
Z}_2$ by using a field theoretical construction. This discrepancy is
crucial when one discusses the reconnection of vortices because the
latter contains an orbifold singularity.  The study of non-Abelian
vortices of higher winding numbers (or equivalently, of  composite
vortices) can be important in the understanding of confinement
mechanism, or in the detailed model study of cosmic string
interactions. Motivated by these considerations,  we study  in this
paper  the moduli space of $k=2$  axially symmetric non-Abelian
vortices of $U(N)$ gauge theories, by using the method of moduli
matrix \cite{Eto:2005yh}, which was originally introduced in the
study of domain walls in \cite{Isozumi:2004jc} (see
\cite{Eto:2006pg} for  a review).

This paper is organized as follows.  In Section \ref{NAV} we briefly
review the non-Abelian vortices and their moduli matrix description
in the context of  most frequently discussed models: $U(N)$  gauge
theory with $N$ flavors of scalar quarks, and discuss, in the case
of $U(2)$ model, the  transformation properties of the fundamental
$k=1$ vortices. The moduli space and the transformation properties
of  $k=2$  co-axial vortices are  studied in Section \ref{u(2)},
where we reproduce  the results by Hashimoto and Tong, and by Auzzi,
Shifman and Yung, and resolve the apparent discrepancy between their
results.  We generalize our results on $k=2$ co-axial vortices to
analogous vortices of $U(N)$  theories  in Section \ref{general}. 
In Appendices A and B 
we present detailed comparison between our results and
those by Hashimoto and Tong, and by Auzzi, Shifman and Yung. 
In Appendix C, we show the relation between 
the K\"ahler quotient construction and 
the moduli matrix approach.
In Appendix D we give a simple method to 
construct the moduli matrix for vortices 
of higher winding number as a product of 
those for fundamental vortices.

\section{Non-Abelian vortices}\label{NAV}

\subsection{Vortex equations}

Our model is  an  $U(N)_{\rm G}$ gauge theory coupled with $N$ Higgs fields
in the fundamental representation,
denoted by an $N$ by $N$ complex (color-flavor) matrix $H$.
The Lagrangian is given by
\beq
{\cal L} = \Tr \left[ - \frac{1}{2g^2} F_{\mu\nu}\, F^{\mu\nu} -
\D_\mu \,H \;\D^\mu  H^\dagger - \lambda \left( c\,{\bf 1}_N -
H\,H^\dagger\right)^2\right]
\eeqq
where $F_{\mu\nu} = \partial_\mu W_\nu - \partial_\nu W_\nu + i \left[W_\mu,W_\nu\right]$
and $\D_\mu H  = \left(\partial_\mu + i\, W_\mu\right)\, H$. 
$g$ is the $U(N)_{\rm G}$ gauge coupling and $\lambda$
is a scalar quartic coupling.
In this article we shall restrict ourselves to
the critical case, $\lambda = \frac{g^2}{4}$ (BPS limit):
 in this case the  model
can be regarded as the bosonic sector of a supersymmetric gauge
theory. In the supersymmetric context the constant $c$ is the
Fayet-Iliopoulos parameter. In the following we set $c>0$ to ensure
stable vortex configurations. This model has an $SU(N)_{\rm F}$
flavor symmetry acting on $H$ from  the right while the $U(N)_{\rm
G}$ gauge symmetry acts on $H$ from the  left. The vacuum of this
model, determined by $H\, H^\dagger = c\,{\bf 1}_N$, is unique up to
a gauge transformation and is in the  Higgs phase. The $U(N)_{\rm
G}$ gauge symmetry is completely broken. This vacuum preserves a
global unbroken symmetry $SU(N)_{\rm G+F}$  (the color-flavor
locking phase).

This system admits
the Abrikosov-Nielsen-Olesen (ANO) type of vortices 
\cite{Abrikosov:1956sx,Nielsen:cs}
which saturate Bogomol'nyi  bound.
The equations of motion reduce to the first order
non-Abelian vortex  equations \cite{Hanany:2003hp,Auzzi:2003fs,Eto:2005yh,Eto:2006pg}:
\beq
\left (\D_1+i\D_2\right) \, H = 0,\quad
F_{12} + \frac{g^2}{2} \left( c \,{\bf 1}_N - H\, H^\dagger\right) =0.
\label{eq:bps_equation}
\eeqq
It turns out that the matter  equation can be solved
by
\beq
 H = S^{-1}(z,\bar z) \, H_0(z),\quad
 W_1 + i\,W_2 = - 2\,i\,S^{-1}(z,\bar z) \, \bar\partial_z S(z,\bar z).
\label{eq:bps_sol}
\eeqq
where  the elements of  the  $N$ by $N$  moduli matrix $H_0(z)$
are  holomorphic functions of the complex coordinate
$z \equiv x^1+ix^2$ \cite{Isozumi:2004vg,Eto:2005yh,Eto:2006pg}, and
 $S$ is an $N$ by $N$ matrix invertible  over the whole $z$-plane.
For any given $H_0(z)$,
$S$ is uniquely determined up to a gauge transformation by
the second equation in Eq.~(\ref{eq:bps_equation}).
The physical fields $H$ and $W$
are obtained by plugging the solution $S$ back into Eq.~(\ref{eq:bps_sol}).
Each element of the matrix $H_0(z)$ must be a polynomial of $z$ in order
to satisfy the boundary condition,  $\det H_{0}(z) = {\cal O}(z^{k})$, 
$k$ being  the vortex (winding) number.

A great advantage of  the method lies in  the fact that all the
integration constants of the BPS equations
Eq.~(\ref{eq:bps_equation}) - moduli parameters - are encoded  in
the moduli matrix $H_0(z)$ as various coefficients of the
polynomials, justifying  its name \cite{Eto:2005yh,Eto:2006pg}. The
zeros $\{z_i\}$ of $$ \det H_0(z)\propto \prod_{i}(z-z_i) $$    can
be interpreted as  the positions of the component  vortices,  when
they are sufficiently  far apart from each other.  Vice versa, when
they overlap significantly,  $z_{i}$'s have no clear physical
meaning as the center of each component vortex: they are just part
of the moduli parameters, characterizing  the shape and color-flavor
orientation of  the vortex under consideration.

Notice that the rank of $H$ gets  reduced  by one at vortex positions $z_i$
when all the vortices are separated, $z_i \neq z_j$ for $i\neq j$.
A constant vector defined by
\beq
 H\big|_{z=z_i}\vec\phi_i = 0
\label{orientation}
\eeq
is associated with each component  vortex at $z=z_i$.
An overall constant of $\vec \phi_i$
cannot be determined from Eq.~(\ref{orientation})
so we should introduce an equivalence relation ``$\sim$", given by
\begin{eqnarray}
 \vec \phi_i\, \sim \, \lambda \, \vec\phi_i, \quad {\rm with~}
\lambda \in {\bf C}^*. \label{eq:CP}
\end{eqnarray}
Thus, each vector $\vec\phi_i$ takes a value in the projective space
${\bf C}P^{N-1} = SU(N)/[SU(N-1) \times U(1)]$. This space can be
understood as a space parameterized by Nambu-Goldstone modes
associated with the symmetry breaking,
\begin{eqnarray}
 SU(N)_{{\rm G}+{\rm F}}\rightarrow U(1)\times SU(N-1),
\end{eqnarray}
caused by the presence of a vortex
\cite{Hanany:2003hp,Auzzi:2003fs,Eto:2004rz,Eto:2005yh,Eto:2006pg}.
We call $\phi_i$  the  orientational vector.

The solutions Eq.~(\ref{eq:bps_sol}) are invariant under
\beq
 (H_0,S) \to (V(z)H_0,V(z)S)  \label{V-trans}
\eeq
with $V(z) \in GL(N,{\bf C})$ being holomorphic with respect to $z$.
We call this   $V$-transformation or $V$-equivalence relation.
The moduli space of the vortex equations Eq.~(\ref{eq:bps_equation})
is obtained as the quotient space
${\cal M}_{\rm total} = \{H_0(z)\}/GL(N,{\bf C})$.
This space is infinite dimensional and
can be decomposed into topological sectors according to the vortex  number $k$.
The $k$-th topological sector ${\cal M}_{N,k}$,
the moduli space of $k$ vortices, is determined by
the condition that $\det H_0(z)$ is of order $z^k$:
\beq
{\cal M}_{N,k} \simeq
\Big\{H_0(z)\Big|\det H_0(z) = {\cal O}(z^k)\Big\} / \{V(z)\}.
\label{eq:MSbyH_0}
\eeqq

\subsection{Fundamental ($k=1$)  vortices } \label{sec:single}
Let us first  discuss a single non-Abelian vortex
in $U(2)$ gauge theory.
The condition on the moduli matrix $H_0$ is
$\det H_0 = {\cal O}(z)$.
Modulo  $V$-equivalence relation Eq.~(\ref{V-trans}),  the moduli matrix can be brought to one of the
 following two forms \cite{Eto:2004rz}:
\beq
 H_{0}^{(1,0)}(z)
  = \left(\begin{array}{cc}z-z_0 & 0 \\ -b'  & 1\end{array}\right), \qquad
 H_{0}^{(0,1)}(z)
  = \left(\begin{array}{cc} 1  &  - b \\ 0 & z-z_0\end{array}\right)
 \label{minimum}
\eeqq with $b,b'$ and $z_0$ complex parameters. Here $z_0$ gives the
position moduli whereas $b$ and $b'$ give the orientational moduli
as we see below. The two matrices in Eq.~(\ref{minimum}) describe
the same single vortex  configuration but in two   different patches
of the moduli space. Let us denote them  ${\cal U}^{(1,0)} =
\{z_0,b' \}$ and ${\cal U}^{(0,1)} = \{z_0,b \}$. The transition
function between these patches is given, except for the point $b'=0$
in the patch ${\cal U}^{(1,0)}$ and $b=0$ in ${\cal U}^{(0,1)}$, by
the $V$-transformation Eq.~(\ref{V-trans})  of the  form
\cite{Eto:2004rz} \beq
 V =
\left(
\begin{matrix}
0 & -1/b' \\
b' & z-z_0
\end{matrix}
\right) \in GL(2,{\bf C}).
\eeq
This yields the transition function
\beq
b= \frac{1}{b'},\qquad
(b,b' \neq 0).
\label{simple}
\eeq
$b'$ and $b$ are seen to be  the two patches of a ${\bf C}P^1$, leading to
the conclusion that the moduli space of the single non-Abelian vortex
is
\beq
{\cal M}_{N=2,k=1} \simeq {\bf C} \times {\bf C}P^1,
\eeq
where the first factor ${\bf C}$ corresponds to the position $z_0$ of
the vortex.

The same conclusion can be reached from the orientation vector,
defined by $H_0(z=z_0) \, \vec \phi = 0$:
 $\vec\phi$ is given by
\beq
\vec \phi \sim
\left(
\begin{array}{c}
1\\
b'
\end{array}
\right)
\sim
\left(
\begin{array}{c}
b\\
1
\end{array}
\right). \label{eq:phi-b} \eeq We see that the components of $\vec
\phi$ are the  homogeneous coordinates of ${\bf C}P^1$; $b,b'$ are
the inhomogeneous coordinates.\footnote{Similarly, in the case of 
$U(N)$ gauge theory, the components of $\vec\phi$ correspond 
to the homogeneous coordinates of ${\bf C}P^{N-1}$.}

The individual vortex breaks the color-flavor diagonal symmetry
$SU(2)_{{\rm G}+\rm F}$, so that it transforms nontrivially under
it.  The transformation property of the vortex moduli parameters can
be conveniently  studied by the $SU(2)_{\rm F}$ flavor
transformations on the moduli matrix, as the color transformations
acting from the left can be regarded as a $V$ transformation. 
The flavor symmetry acts on $H_{0}$  as $H_0 \to H_0 \,U$ with $U
\in SU(2)_{\rm F}$. A general $SU(2)$ matrix
\beq U = \left(
\begin{matrix}
u & v\\
-v^* & u^*
\end{matrix}
\right) \label{su2}, \eeq with $u,v \in {\bf C}$ satisfying
$|u|^2+|v|^2=1$, acts for instance on $H_0^{(0,1)}$ in
Eq.~(\ref{minimum}) as
\begin{eqnarray}
 H_{0}^{(0,1)}(z)&\to& H_{0}^{(0,1)}(z) \, U
=\left(\begin{array}{cc} u+v^* \,b  &  -u^*\,  b+v \\
-v^*\,(z-z_0)& u^*\,(z-z_0)
\end{array}\right).
\eeq
The right hand side should be
pulled back to the form $H^{(0,1)}_0$ in Eq.~(\ref{minimum})
by using an appropriate  $V$-transformation Eq.~(\ref{V-trans}).
This can be achieved by
\beq
V_U\, H_{0}^{(0,1)}(z)\, U
=\left(\begin{array}{cc}
1  &  -\frac{u^* \,b - v}{v^*\, b +u}\\
0 & z-z_0\end{array}\right),\quad V_U = \left(
\begin{matrix}
(u+v^*\,b)^{-1} & 0\\
v^*\,(z-z_0) & u+v^*\,b
\end{matrix}
\right)\in GL(2,{\bf C}) .
\end{eqnarray}
The $SU(2)_{\rm G+F}$ transformation law of $b$  is then
\begin{eqnarray}
 b\rightarrow \frac{u^* \, b-v}{v^* \, b+u}, \label{eq:su(2)_k1}
\end{eqnarray}
which  is the standard $SU(2)$ transformation law of the
inhomogeneous coordinate of ${\bf C}P^1$.\footnote{ The coordinate
$b$ is invariant (more precisely the orientational vector receives a
global phase) under the $U(1)$ transformations generated by 
$\hat{\bf n}\cdot \vec{\bf \sigma}/2$, $\hat {\bf n}= {1\over (1+|b|^2)}\;(2\, \Re\, b
,2\,\Im\, b , |b|^2-1)$. This implies the coset structure ${\bf
C}P^1 \simeq \frac{SU(2)}{U(1)}$.\label{u1} }

In terms of the orientational vector $\vec\phi$ in
Eq.~(\ref{eq:phi-b}), the transformation law Eq.~(\ref{eq:su(2)_k1})
can be derived  more straightforwardly.
According to the definition Eq.~(\ref{orientation}),
$\vec\phi$ is transformed in
the fundamental representation of $SU(2)_{\rm F}$:
\beq
\vec\phi \to U^\dagger \vec\phi,\qquad
U \in SU(2)_{\rm F}.
\eeq

\section{$k=2$  Vortices in $U(2)$ Gauge Theory\label{u(2)}}

\subsection{Moduli space of $k=2$ vortices}

Configurations of $k=2$ vortices at arbitrary positions are given by
the moduli matrix whose determinant has degree two, $\det H_0 =
{\cal O}(z^2)$. By using $V$-transformations Eq.~(\ref{V-trans}) the
moduli matrix satisfying this condition can be brought  into one of
the following three forms \cite{Eto:2005yh,Eto:2006pg}
\begin{eqnarray}
&H_0^{(2,0)} = \left(
\begin{array}{cc}
z^2 - \alpha' \, z - \beta' & 0\\
-a'\, z- b' & 1
\end{array}
\right),\
H_0^{(1,1)} =
\left(
\begin{array}{cc}
z-\phi & -\eta\\
-\tilde\eta & z-\tilde\phi
\end{array}
\right), \nonumber\\
&H_0^{(0,2)}= \left(
\begin{array}{cc}
1 & -a\,z-b\\
0 & z^2-\alpha \,z - \beta
\end{array}
\right).
   \label{eq:HofN2k2}
\end{eqnarray}
These
define the three patches
${\cal U}^{(2,0)}=\{a',b',\alpha',\beta'\}$,
${\cal U}^{(1,1)}=\{\phi,\tilde\phi,\eta,\tilde\eta\}$,
${\cal U}^{(0,2)}=\{a,b,\alpha,\beta\}$
of  the moduli space ${\cal M}_{N=2,k=2}$.
The transition from ${\cal U}^{(1,1)}$ to ${\cal U}^{(0,2)}$
is given via the $V$-transformation
$
V =
\left(
\begin{smallmatrix}
0 & -1/\tilde\eta\\
\tilde\eta & z - \phi
\end{smallmatrix}
\right)$:
\begin{eqnarray}
 a=\frac{1}{ \tilde \eta },\quad
 b= -  \frac{\tilde \phi }{ \tilde \eta },\quad
\alpha = \phi + \tilde \phi ,\quad \beta =\eta \, \tilde \eta -\phi
\,\tilde \phi. \label{eq:02to11}
\end{eqnarray}
Similarly  the transition from ${\cal U}^{(2,0)}$ to
${\cal U}^{(0,2)}$ is given by
$
V =
\left(
\begin{smallmatrix}
\frac{-{a'}^2}{{a'}^2\,\beta' - a'\,b'\,\alpha' -{b'}^2} &
\frac{-a'\,z + a'\,\alpha' + b'}{{a'}^2\,\beta' - a'\,b'\,\alpha' -{b'}^2} \\
a'\,z + b' & z^2 - \alpha'\,z - \beta'
\end{smallmatrix}
\right)$,
which yields
\begin{eqnarray}
 a= {a'\over {a'}^2\, \beta' -a'\, b'\, \alpha' -{b'}^2},\quad
 b=-{b'+a'\, \alpha' \over {a'}^2\, \beta' -a'\,b'\,\alpha' -{b'}^2},\quad
 \alpha = \alpha',\quad \beta = \beta'.
\label{eq:02to20}
\end{eqnarray}
Finally those between
${\cal U}^{(1,1)}$ and
${\cal U}^{(2,0)}$ are given by the
composition of  the  transformations   Eq.~(\ref{eq:02to11}) and Eq.~(\ref{eq:02to20}).
Let us now discuss the moduli space of $k=2$   vortices
separately for the cases  where  the two vortex centers  are 1) distinct  ($z_{1}\ne z_{2}$),  and
2) coincident $(z_{1}=z_{2})$.

1)   At the vortex positions $z_i$ the orientational vectors are determined by
Eq.~(\ref{orientation}).
The orientational vectors $\vec\phi_i$ $(i=1,2)$ are then obtained by
\begin{eqnarray}
 \vec \phi _i\sim
\left(\begin{array}{c}
a z_i+b\\ 1  \end{array}\right)
\sim  \left(\begin{array}{c}
z_i-\tilde \phi \\ \tilde \eta   \end{array}\right)
\sim
\left(\begin{array}{c}
\eta \\  z_i-\phi   \end{array}\right)
\sim
\left(\begin{array}{c}
1\\a' z_i+b'  \end{array}\right).
   \label{eq:vtx:fourkinds}
\end{eqnarray}
The two $(i=1,2)$ parameters defined by $b_i\equiv a \, z_i+b$ (or
$b_i'\equiv a' \, z_i+b'$) parameterize the two different ${\bf
C}P^1$'s separately. Conversely if the two vortices are separated
$z_1\neq z_2$, then moduli parameters $a, b$ are described by $b_1,
b_2$ with positions of vortices $z_1,z_2$ as
\begin{eqnarray}
 a=\frac{b_1-b_2}{z_1-z_2}, \quad b=\frac{b_2 \, z_1-b_1 \, z_2}{z_1-z_2},
\quad \alpha=z_1+z_2, \quad \beta=-z_1 \, z_2, \label{eq:ab-b_12}
\end{eqnarray}
(and similar relations for the primed variables). Thus in the case
of separated vortices $\{b_1, b_2, z_1, z_2\}$
($\{b_1',b_2',z_1,z_2\}$) can be taken as appropriate coordinates of
the moduli space, instead of ${\{a,b,\alpha,\beta\}}$
(${\{a',b',\alpha',\beta'\}}$). The transition functions are also
obtained by applying the equivalence relation  Eq.~(\ref{eq:CP}) to
 Eq.~(\ref{eq:vtx:fourkinds}), for instance,
\begin{eqnarray}
 b_i=\frac{1}{b'}_i \quad (b_i,b'_i\neq 0) .\label{eq:bb'2}
\end{eqnarray}
It can be shown  that these are equivalent to Eqs.~(\ref{eq:02to20})
by use of  Eq.~(\ref{eq:ab-b_12}) and analogous relation for the
primed parameters. The coordinates in ${\cal U}^{(1,1)}$ are also
the orientational moduli. If we take $\{b_1,b_2',z_1,z_2\}$ as a set
of independent moduli and substitute Eq.~(\ref{eq:ab-b_12}) and
$b_2=1/b_2'$ to Eq.~(\ref{eq:02to11}), then we obtain, for $b_1 \,
b_2'\neq 1$
\begin{eqnarray}
 \phi= \frac{z_2-b_1 \, b_2'z_1}{1-b_1 \,b_2'},
\quad\eta=\frac{z_1-z_2}{1-b_1\,b_2'} \, b_1, \quad \tilde \phi=
\frac{z_1-b_1\,b_2'\,z_2}{1-b_1\,b_2'},\quad \tilde
\eta=-\frac{z_1-z_2}{1-b_1\,b_2'}\,b_2'. \label{eq:phis-b_12}
\end{eqnarray}

It can be seen that  the representation  Eq.~(\ref{eq:ab-b_12})
implies that ${\cal U}^{(0,2)}$ and ${\cal U}^{(2,0)}$ are suitable
for describing the situation when two orientational moduli
are parallel or   nearby.
On the other hand,
Eq.~(\ref{eq:phis-b_12}) implies that ${\cal U}^{(1,1)}$ is suitable to
describe the situation when orientational moduli are
orthogonal or close to  such a situation, while
not adequate for describing  a parallel set.
Therefore,  the moduli space for two separated vortices are completely described by
the positions and the two orientational moduli   $b_1,b_2, \, (b_1',b_2')$:
 the moduli space for the composite vortices  in this case  is given by
\cite{Hashimoto:2005hi,Eto:2005yh} \beq {\cal M}_{k=2,N=2}^{\rm
separated} \simeq \left( {\bf C} \times {\bf C}P^1\right)^2/{\mathfrak S}_2,
\eeq where ${\mathfrak S}_2$  permutes  the centers and orientations   of the
two vortices.

\medskip
2)
We now focus on coincident (co-axial) vortices
($z_1 = z_2$), with the moduli space denoted by
\beq
 \tilde {\cal M}_{N=2,k=2} \equiv {\cal M}_{N=2,k=2}\big|_{z_1=z_2}.
\eeq
As an overall translational moduli is trivial,
 we  set $z_1 = z_2 = 0$ without loss of generality.
According to Eqs.~(\ref{eq:ab-b_12}) and Eq.~(\ref{eq:phis-b_12}),
all points in the moduli space tend to
the origin of ${\cal U}^{(1,1)}$
in the limit of $z_2\rightarrow z_1$,
as long as $b_1$ and $b_2$ take different values.
 A more careful treatment is needed
in this case.
In terms of the moduli matrix, the condition
of coincidence is given by
$
 \det H_0(z) = z^2
$.
We have
\beq
 \left\{ \alpha = 0,\ \beta =0 \right\},\quad
 \left\{ \tilde \phi = - \phi,\  \phi \, \tilde\phi - \eta \, \tilde\eta = 0 \right\}
 \quad {\rm and} \quad
 \left\{ \alpha' = 0,\ \beta' = 0 \right\},
 \label{eq:consident-cond}
\eeq
in  ${\cal U}^{(2,0)}$, ${\cal U}^{(1,1)}$ and ${\cal U}^{(0,2)}$,
respectively.
  $\tilde {\cal M}_{N=2,k=2}$
is covered by the reduced patches $\tilde {\cal U}^{(2,0)}$,
$\tilde {\cal U}^{(1,1)}$ and $\tilde {\cal U}^{(0,2)}$, defined by
the moduli matrices 
\beq
 H_0^{(2,0)}=
 \left(
 \begin{array}{cc}
       z^2 & 0\\
  -a'\,z- b' & 1
 \end{array}
 \right), \
 H_0^{(1,1)}=
\left(
\begin{array}{cc}
         z-\phi & -\eta\\
 -\tilde\eta & z+\phi
\end{array}
\right),\
 H_0^{(0,2)}=
 \left(
\begin{array}{cc}
 1 & -a\, z-b\\
 0 & z^2
\end{array}
\right).
\label{eq:HofN2k2-2}
\eeqq
The following constraint exists among the coordinates in
$\tilde {\cal U}^{(1,1)}$:
\beq
 \phi^2 + \eta \, \tilde\eta = 0. \label{eq:constraint}
\eeqq
The transition functions between $\tilde {\cal U}^{(0,2)}$ and
$\tilde {\cal U}^{(1,1)}$ are given by
\begin{eqnarray}
a={1\over \tilde \eta },\quad
b= {\phi \over \tilde \eta } = - \frac{\eta}{\phi},
\label{eq:02to11_co}
\end{eqnarray}
and those between  $\tilde {\cal U}^{(0,2)}$ and
$\tilde {\cal U}^{(2,0)}$  by
\begin{eqnarray}
 a= - {a'\over {b'}^2},\quad
 b= {1 \over {b'}}.
 \label{eq:02to20_co}
\end{eqnarray}

All the patches defined by Eq.~(\ref{eq:HofN2k2-2}) are parameterized
by two independent complex parameters. The reduced patches $\tilde
{\cal U}^{(2,0)}$ and $\tilde {\cal U}^{(0,2)}$ are  locally
isomorphic to ${\bf C}^2$   with $a,b$ or $a^{\prime}, b^{\prime} $
being good coordinates. However, $\tilde {\cal U}^{(1,1)}$  suffers
from the constraint Eq.~(\ref{eq:constraint}) which gives the
$A_1$-type (${\bf Z}_2$) orbifold singularity at the origin and
therefore $\tilde {\cal U}^{(1,1)} \simeq {\bf C}^2/{\bf Z}_2$
locally (See Eq.~(\ref{eq:Z2-ident}), below). Note that the moduli
matrix $H_0(z)$ is proportional to the unit matrix at the
singularity: $H_0(z)=z  \, {\bf 1}_2$. This implies that
configurations of the physical fields ($H$ and $F_{12}$) are also
proportional to the unit matrix where the global symmetry
$SU(2)_{{\rm G}+{\rm F}}$ is fully recovered  at that singularity.
The full gauge symmetry is also recovered at the core of coincident
vortices.

{\small {\bf  Remark:} A brief comment on the  orientational
vectors. We could  extract a part of moduli in the moduli matrix as
the orientational vector at $z = 0$, as in  the case of separated
vortices discussed above: \beq \vec\phi \sim \left(
\begin{array}{c}
1 \\
b'
\end{array}
\right)
\sim
\left(
\begin{array}{c}
\eta \\
- \phi
\end{array}
\right)
\sim
\left(
\begin{array}{c}
\phi \\
\tilde\eta
\end{array}
\right)
\sim
\left(
\begin{array}{c}
b \\
1
\end{array}
\right).
\label{eq:phi_k2N2_co}
\eeq
From the identification $\vec\phi \sim \lambda \, \vec\phi\ (\lambda
\in {\bf C}^*)$ with the transition functions given in
Eqs.~(\ref{eq:02to11_co}) and Eq.~(\ref{eq:02to20_co}), we find that
the orientational moduli again parameterizes ${\bf C}P^1$. However,
the orientational vectors in Eq.~(\ref{eq:phi_k2N2_co}) are not
sufficient  to pick up all the moduli parameters in the moduli
matrix $H_0$. For instance $a$ is lost  in the $\tilde {\cal
U}^{(0,2)}$ patch.
 It is even  ill-defined
at the singular point, as  $H_0^{(1,1)}(z=0) = 0$.  }

\begin{table}[ht]
\begin{center}
\ \\
\begin{tabular}{c|ccc}
\setlength{\baselineskip}{7.5\baselineskip}
      &  $(a,b)$ & $(a',b')$ & $(X,Y)$ \\ \hline
 \rule{0pt}{4ex} $(a,b) =$  & ** & $(-{a' / b'{}^2}, {1 / b'})$
               & $(-{1/ Y^2},  {X / Y})$ \\
 \rule{0pt}{4ex} $(a',b')=$ & $({- a / b^2},{1/ b})$ & **
               & $({1 / X^2},{Y/ X})$ \\
 \rule{0pt}{4ex} $(X,Y) =$ & $( \pm i {b / \sqrt a}, \pm i {1/ \sqrt a})$
            & $( \pm {1/ \sqrt{a'}}, \pm  {b' / \sqrt{a'}})$ & **

\end{tabular}
\caption{\small Transition functions between the three patches
$\tilde {\cal U}^{(2,0)}$, $\tilde {\cal U}^{(1,1)}$ and $\tilde {\cal
U}^{(0,2)}$.} \label{table:transition function}
\end{center}
\end{table}
To clarify the whole structure of the space
$\tilde {\cal M}_{N=2,k=2}$,
let us define new coordinates,  solving  the constraint
Eq.~(\ref{eq:constraint})
\beq
 X \, Y \equiv - \phi, \quad
 X^2 \equiv \eta, \quad
 Y^2 \equiv -\tilde \eta.
\eeqq
This clarifies the structure of the singularity at the origin.
The coordinates $(X,Y)$ describe the patch $\tilde {\cal
U}^{(1,1)}$  correctly  {\it modulo}   ${\bf Z}_2$
identification
\beq
 (X,Y) \sim (-X,-Y) .
 \label{eq:Z2-ident}
\eeqq Using the transition functions Eq.~(\ref{eq:02to11_co}) and
Eq.~(\ref{eq:02to20_co}), the three local  domains are patched
together   as in Table \ref{table:transition function}. In terms of
the new coordinates $(X,Y)$, the orientational vector defined at
$z=0$ is given by \beq \vec \phi \sim \left(
\begin{array}{c}
1 \\
b'
\end{array}
\right)
\sim
\left(
\begin{array}{c}
X \\
Y
\end{array}
\right)
\sim
\left(
\begin{array}{c}
b \\
1
\end{array}
\right)
\label{eq:phi_k2N2_co2}
\eeq
with $\vec\phi \sim \lambda \, \vec\phi\ (\lambda \in {\bf C}^*)$.
This equivalence relation recovers the transition functions between
$b,b'$ and $(X,Y)$ in the Table \ref{table:transition function}.
These are coordinates on the ${\bf C}P^1$  as was mentioned above.
But  this ${\bf C}P^1$ is only a subspace of the moduli space
$\tilde {\cal M}_{N=2,k=2}$.

The full space $\tilde {\cal M}_{N=2,k=2}$ can be made visible by
attaching the remaining parameters $a,a'$ to ${\bf C}P^1$. We
arrange the moduli parameters in the three  patches $\tilde {\cal
U}^{(2,0)}$, $\tilde {\cal U}^{(1,1)}$ and $\tilde {\cal U}^{(0,2)}$
as \beq \left(
\begin{array}{c}
a'\\
1\\
b'
\end{array}
\right)
\sim
\left(
\begin{array}{c}
1\\
X\\
Y
\end{array}
\right)
\sim
\left(
\begin{array}{c}
-a\\
b\\
1
\end{array}
\right),
\label{eq:3patch_wcp2}
\eeq
respectively,
with the  equivalence relation ``$\sim$", {\it defined}  by
\beq
\left(
\begin{array}{c}
\phi_0\\
\phi_1\\
\phi_2
\end{array}
\right)
 \sim
\left(
\begin{array}{c}
\lambda^2\, \phi_0\\
\lambda\, \phi_1\\
\lambda \, \phi_2
\end{array}
\right) \qquad {\rm with}\quad
 \lam \in {\bf C}^* .
 \label{eq:equiv}
\eeqq
All the transition functions in Table \ref{table:transition function}
are then nicely reproduced.
The equivalence relation Eq.~(\ref{eq:equiv}) defines
a {\it weighted complex projective space}
with the weights $(2,1,1)$.
We thus conclude that
the moduli space for the coincident (coaxial)
$k=2$   non-Abelian vortices is
a weighted projective space,
\beq
 \tilde {\cal M}_{N=2,k=2} \simeq W{\bf C}P^2_{(2,1,1)}.
  \label{eq:w-CP2}
\eeqq

While the complex projective spaces with common weights, ${\bf
C}P^n$, are smooth, weighted projective spaces have singularities.
In fact, we have shown that  $\tilde {\cal U}^{(1,1)} \simeq {\bf
C}^2/{\bf Z}_2$, and it has a conical singularity at the origin by
$(1,X,Y) \sim (1,-X,-Y)$, whose existence  was first pointed out
by ASY \cite{Auzzi:2005gr}. The origin of the conical singularity
can be seen clearly from the equivalence relation
Eq.~(\ref{eq:equiv}). As mentioned above the transition functions in
Table \ref{table:transition function} are reproduced via the
equivalence relation Eq.~(\ref{eq:equiv}). In fact, one finds that
$\lam = {1 \over X}$ gives $(\lambda ^2,\lambda \, X,\lambda \, Y) =
(a',1,b')$ and $\lam = {1 \over Y}$ gives $(\lambda^2 ,\lambda \,
X,\lambda \, Y) = (-a,b,1)$. Note that $\lambda$ in the equivalence
relation Eq.~(\ref{eq:equiv}) is completely fixed in the patches
$\tilde {\cal U}^{(2,0)}$ and $\tilde {\cal U}^{(0,2)}$ given in
Eq.~(\ref{eq:3patch_wcp2}). However, in the middle patch $(1,X,Y)$
we still have a freedom $\lambda = -1$ which leaves  the first
component $1$ untouched, but  changes $(1,X,Y) \to (1,- X,-Y)$.

The relation between  our result and that in \cite{Auzzi:2005gr}
becomes clear by defining $\xi^2 \equiv \phi_0$ ($\xi = \pm
\sqrt{\phi_0}$). Now  the parameters $(\xi,\phi_1,\phi_2)$ have a
common weight $\lambda$, so they can be regarded  as the homogeneous
coordinates  of ${\bf C}P^2$. But one  must identify $\xi \sim
-\xi$ clearly, and this leads to the ${\bf Z}_2$ quotient
$(\xi,\phi_1,\phi_2) \sim (\xi, - \phi_1, - \phi_2)$. Therefore
our moduli space can also be rewritten as \beq \tilde {\cal
M}_{N=2,k=2} \simeq {\bf C}P^2/{\bf Z}_2 \eeqq reproducing the
result of \cite{Auzzi:2005gr}. 
Such a ${\bf Z}_2$ equivalence, however, does not change 
the homotopy of
$ {\cal M}_{N=2,k=2}$: it remains $ {\bf C}P^2$
\cite{Hashimoto:2005hi}. This is analogous to an $(x,y)\sim(-x, -y)$
equivalence relation (with real $x,y$) introduced in one local
coordinate system of ${\bf C}P^1$ (a sphere), which leads to a
sphere with two conic singularities (a rugby ball, or a lemon)
instead of the original smooth sphere. \footnote{For instance, it is
easily seen that ${\cal M}_{N=2,k=2}\simeq {\bf C}P^2/{\bf Z}_2$
remains simply connected. The higher homotopy groups cannot change
by a discrete fibration \cite{Dubrovin}.}
See Appendices \ref{ASY} and \ref{HTASY}  for more details.

\subsection{$SU(2)$ transformation law of co-axial $k=2 $  vortices}

The complex projective space
${\bf C}P^2 \simeq \frac{SU(3)}{SU(2)\times U(1)}$
with the Fubini-Study metric has an $SU(3)$ isometry.
On the other hand, the weighted projective space
$W{\bf C}P^2_{(2,1,1)}$ can have an $SU(2)$ isometry at most
due to the difference of the weights Eq.~(\ref{eq:equiv}).
This matches with the fact that
we have only $SU(2)_{\rm G+F}$ symmetry
acting on the moduli space.
In this subsection we investigate
the $SU(2)_{\rm G+F}$ transformation laws of
the moduli for the
co-axial two vortices,
as was done for the fundamental  vortex in section \ref{sec:single}.

Let us start with the patch $\tilde {\cal U}^{(0,2)}$,  with
the moduli matrix
\beq
H_0^{(0,2)} =
\left(
\begin{array}{cc}
1 & - a \, z - b\\
0 & z^2
\end{array}
\right)
  \label{H02}
\eeq
An $SU(2)$ matrix $ U $ like Eq.~(\ref{su2}) acts on the above
$H_0$ from the right, \beq H_0^{(0,2)} \to 
H_0^{(0,2)} \, U = \left(
\begin{array}{cc}
v^* \, a \, z + u + v^* \, b & -u^* \, a \, z + v - u^* \, b\\
-v^* \, z^2 & u^* \, z^2
\end{array}
\right).
\eeq
A $V$-transformation $V= V_1\, V_2$, where
\beq
V_1 =
\left(
\begin{matrix}
- \frac{v^* \, a}{\left(u + v^* \, b\right)^2} & 0 \\
0 & 1
\end{matrix}
\right), \quad
V_2 =
\left(
\begin{matrix}
z - \frac{u + v^* \, b}{v^* \, a} & a\\
v^* \, z^2 & v^* \, a \, z + u + v^* \, b
\end{matrix}
\right)
\eeq
brings the result back to the upper right triangle form,
\beq
H_0^{(0,2)} \to
H_0^{(0,2)} \, U
\sim
V \, H_0^{(0,2)} \, U
= \left(
\begin{array}{cc}
1 & - \frac{a}{\left(u + v^* \, b\right)^2} z
+ \frac{v - u^* \, b}{u + v^* \, b}\\
0 & z^2
\end{array}
\right). \eeq The  $SU(2)$ transformation laws of the  parameters
$a,b$ are then
\beq a \to  \frac{a}{\left(v^* \, b + u\right)^2},\qquad
b \to  \frac{u^* \, b - v}{v^* \, b + u}. \label{eq:tcp1}
\eeq
As in the previous section, $b$ can be  regarded as an inhomogeneous
coordinate of ${\bf C}P^1$; in fact, $b$ is invariant under a $U(1)$
subgroup (see footnote \ref{u1}) and this means that $b$
parameterizes ${\bf C}P^1 \simeq \frac{SU(2)}{U(1)}$.
 On the other hand,
the transformation law of the parameter $a$ can be re-written  as
\beq a \to \left[ \frac{d}{db} \left(\frac{u^* \, b - v}{v^* \,b +
u}\right) \right] a,
\eeq
showing  that the parameter $a$ is a tangent vector on the base
space ${\bf C}P^1$ parameterized by $b$. This is very natural. First
recall the situation for separated vortices. The moduli parameters
are extracted from their positions and their orientations,  defined
at the vortex centers.   However, once the vortices overlap exactly
($z_{i}\to z_{j}$), the positions and the orientations only do not
have enough information. 
When $l (\leq k)$ vortices are coincident, 
we need $1,2,\cdots,l-1$ derivatives at the coincident point 
in order to extract all the information. 
They define how vortices 
approach each other $(b_i\to b_j)$ \cite{Eto:2005yh}.

Some  $SU(2)$ action  sends the points in the patch $\tilde {\cal
U}^{(0,2)}$   to where a better  description is in the patch $\tilde
{\cal U}^{(2,0)}$, and vice versa.  Compare Eq.~(\ref{eq:tcp1}) with
$u=0$, $v=i$, with Eq.~(\ref{eq:02to20_co}). This shows indeed that
\beq
\tilde{\cal U}^{(0,2)}\, \cup \,\tilde{\cal U}^{(2,0)}\simeq T{\bf
C}P^1.
\eeqq

\begin{figure}[ht]
\begin{center}
\includegraphics[width=5.5in]{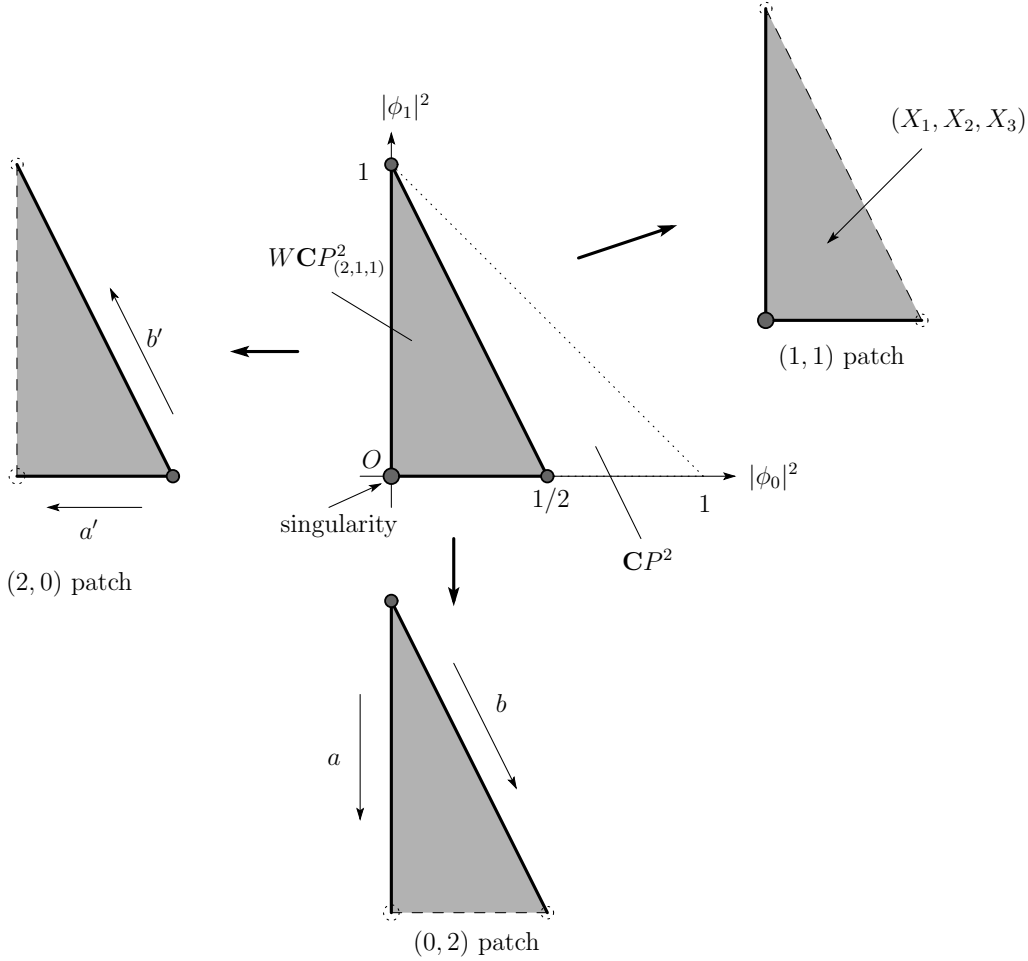}
\caption{Toric diagram of $W{\bf C}P^2_{(2,1,1)}$ and their three patches
$\tilde{\cal U}^{(2,0)}$, $\tilde{\cal U}^{(1,1)}$ and $\tilde{\cal U}^{(0,2)}$.
The diagram is drawn under a gauge fixing condition 
(called the $D$ term constraint)
$\sum_{a=0}^2q^a|\phi_a|^2 = 1$ where $U(1)^{\bf C}$ charges are
$q_a = (2,1,1)$ for $W{\bf C}P^2_{(2,1,1)}$ while $q_a = (1,1,1)$ for
the ordinary ${\bf C}P^2$. 
The triangle with the broken line and $O$ (without singularity) 
denotes the ordinary ${\bf C}P^2$.
}
\label{wcp2}
\end{center}
\end{figure}

Next consider the  patch $\tilde {\cal U}^{(1,1)}$ with
\beq
H_0^{(1,1)} =
\left(
\begin{array}{cc}
z - \phi & -\eta \\
-\tilde\eta & z + \phi
\end{array}
\right),
\qquad
\phi^2 + \eta\, \tilde\eta = 0.
\eeq
It is convenient to rewrite this as
\begin{eqnarray}
 H_0^{(1,1)}=z \, {\bf 1}_2-\vec X\cdot \vec \sigma
\label{eq:11pr}
\end{eqnarray}
where $\vec \sigma$  are  the Pauli matrices and
\beq
\phi \equiv  X_3, \quad
\eta \equiv  X_1 - iX_2,\quad
\tilde \eta \equiv  X_1 + iX_2.
\eeq
$X_1,X_2,X_3$ are then complex coordinates  with a constraint $
X_1^2 + X_2^2 + X_3^2  = 0. $ To keep the form Eq.~(\ref{eq:11pr})
under $SU(2)_{\rm F}$ transformation, we perform  the
$V$-transformation Eq.~(\ref{V-trans}) with $V=U^\dagger$:
$H_0^{(1,1)}\rightarrow U^\dagger \, H_0^{(1,1)} \,U$.
Equivalently, we study the transformation property of the vortex
under $SU(2)_{\rm G + \rm F}$. We find
\beq
 \vec X \cdot \vec \sigma \to
 U^\dagger \left( \vec X \cdot \vec \sigma\right) U,
 \label{eq:adjoint} 
\eeq 
that is, the vector $\vec X$ transforms as
an adjoint (triplet)  representation, except at  $\vec X=0$.   This
last point -  singular point of $W{\bf C}P^2_{(2,1,1)}$ - or the
origin of the patch $\tilde {\cal U}^{(1,1)}$,    is a fixed point
of $SU(2)$ (a singlet). Note also that the transition functions
between the patches $\tilde {\cal U}^{(0,2)}$ and $\tilde {\cal
U}^{(1,1)}$ are given by \beq
 X_3 = \frac{b}{a},\quad
 X_1 - iX_2 = -\frac{b^2}{a},\quad
 X_1 + iX_2 = \frac{1}{a}.
\label{eq:ts_func} \eeq The patch ${\cal U}^{(1,1)}$ does not cover
points at ``infinity", namely the subspace defined by $a = 0$ in the
patch ${\cal U}^{(0,2)}$. That submanifold is nothing but ${\bf
C}P^1$ parameterized by $b$ which is an edge of $W{\bf
C}P^2_{(2,1,1)}$.   See   Fig.~\ref{wcp2}. One can verify that the
transformation law for $a,b$ in Eq.~(\ref{eq:tcp1}) and that for
$\phi,\eta,\tilde\eta$ in Eq.~(\ref{eq:adjoint}) are consistent
through the transition function Eq.~(\ref{eq:ts_func}).  These
results confirm those in \cite{Auzzi:2005gr}.

\section{$k=2$  Vortices in $U(N)$ Gauge Theory} \label{general}

In this section the composition of two non-Abelian vortices  in  a
$U(N)$ gauge theory is systematically investigated.  Up to now we
made use of the direct form of the moduli matrix $H_0(z)$ for
studying the moduli space structure. Another method for studying the
latter will be developed and used to determine the moduli space
below.

\subsection{The case of $U(N)$}
Let ${\bf Z}$ and ${\bf \Psi}$ be
$k$ by $k$ and
$N$ by $k$ constant complex matrices, respectively.
We consider the $GL(k,{\bf C})$ action defined by
\beq
 {\bf Z} \to {\cal V}\, {\bf Z}\,  {\cal V}^{-1}, \quad
 {\bf \Psi} \to {\bf \Psi}\, {\cal V}^{-1} , \quad {\cal V} \in GL(k,{\bf C}).
\eeq It was shown in \cite{Eto:2006pg} that the moduli space ${\cal
M}_{N,k}$ of $k$ vortices can be written as the K\"ahler quotient
\cite{Hitchin:1986ea} defined by \beq {\cal M}_{N,k} \simeq
\left\{{\bf Z},{\bf \Psi}\right\}/\!\!/ GL(k,{\bf C}),
\label{eq:MSbyZPsi} \eeqq where $GL(k,{\bf C})$ action is free on
these matrices.\footnote{ The relation between the moduli matrix
$H_0(z)$ and the two matrices $({\bf Z},{\bf \Psi})$ is given by the
following ADHM-like equation \beq \nabla^\dagger L = 0,\quad \det(z
- {\bf Z}) = \det H_0(z), \eeqq where $ L^\dagger \equiv \left(
H_0(z), {\bf J}(z) \right)$ and $\nabla^\dagger \equiv \left(- {\bf
\Psi}^\dagger, \bar z - {\bf Z}^\dagger \right)$. Here ${\bf J}(z)$
is $N$ by $k$ matrix whose elements are holomorphic function of
$z$.
 $({\bf Z},{\bf \Psi})$ and ${\bf J}(z)$
can be  uniquely determined from a given $H_0(z)$~\cite{Eto:2006pg}.
} The moduli space   ${\cal M}_{N,k}$ given by the moduli matrix
$H_0(z)$ in Eq.~(\ref{eq:MSbyH_0}) and hence  by the complex
K\"ahler quotient in Eq.~(\ref{eq:MSbyZPsi}) is identical to that
obtained by use of the D-brane construction by Hanany-Tong
\cite{Hanany:2003hp}. The concrete correspondence  between them is
obtained by fixing the imaginary part of $GL(k,{\bf C})$ in
Eq.~(\ref{eq:MSbyZPsi}) by the moment map $[{\bf Z}^\dagger,{\bf Z}]
  + {\bf \Psi}^\dagger {\bf \Psi}$:
\beq
 {\cal M}_{N,k}
  \simeq \{ \, ({\bf Z},{\bf \Psi}) \, | \, [{\bf Z}^\dagger,{\bf Z}]
  + {\bf \Psi}^\dagger {\bf \Psi} \propto{\bf 1}_k \, \} \, / \, U(k),
\label{eq:Kahler-quotient1}
\eeqq
where ${\bf Z}$ and ${\bf \Psi}$ are again in the  adjoint and fundamental representations
of $U(k)$ group, respectively.

We now use the
K\"ahler quotient construction
to generalize the discussion to general $N$.
The authors in \cite{Hashimoto:2005hi}
and \cite{Auzzi:2005gr}
used the 
expression Eq.~(\ref{eq:Kahler-quotient1})
but Eq.~(\ref{eq:MSbyZPsi})  
is easier to deal with.
Let us discuss the moduli space of   co-axial  vortices
in terms of Eq.~(\ref{eq:MSbyZPsi}). 
A subspace of the moduli space ${\cal M}_{N,k}$
for coincident vortices at the origin of the $x^1$-$x^2$ plane
is given by putting  the constraint $\det(z-{\bf Z})=z^k$,
that is,
\begin{eqnarray}
 {\rm Tr} \, ({\bf Z}^n)=0, \quad {\rm for~} n=1,2,\cdots k.
\label{eq:const-for-k}
\end{eqnarray}
To understand the subspace clearly
we need to solve the above constraint
by taking appropriate coordinates with $k^2-k$ complex parameters.


In the case of $k=2$, $U(N)$  vortices,
the constraints Eq.~(\ref{eq:const-for-k}) are equivalent to the constraint
Eq.~(\ref{eq:constraint}) for  the $N=2$ case.
 ${\bf Z}$ can be solved in this  case as
\beq
 {\bf Z} = \epsilon \, v \, v^T , \quad
 \epsilon =
 \left(
 \begin{array}{cc}
   0 & 1 \cr
  -1 & 0
 \end{array} \right),
\eeqq
with $v$ a column two-vector with complex components.
The fact that
 the above form of ${\bf Z}$ transforms as  in the   adjoint representation
under $SL(2,{\bf C})\subset GL(2,{\bf C})$
means that $v$ is a fundamental representation of $SL(2,{\bf C})$
since $\bf 2$ and $\bf 2^*$ is equivalent in the $k=2$ case.
Let us define a complex $2$ by $N+1$ matrix by
\beq
 M = \left(\begin{array}{cc}
  {\bf \Psi}^T & v
 \end{array}
 \right).
\label{eq:defM}
\eeqq
The $GL(2,{\bf C}) =  SL(2,{\bf C}) \times {\bf C}^*$
action with elements
${\cal S} \in SL(2,{\bf C})$ and $\lam \in {\bf C}^*$ read
\beq
 M \to {\cal S}\, M , \quad
 \left(\begin{array}{cc}
  {\bf \Psi}^T & v
 \end{array}
 \right) \to
 \left(\begin{array}{cc}
  \lam \, {\bf \Psi}^T & v
 \end{array}
 \right). \label{glquot}
\eeqq The quotient by $GL(2,{\bf C})$ results in a kind of complex
Grassmannian manifold whose ${\bf C}^*$ action has weights
$(\underbrace{1,\cdots,1}_N,0)$. The moduli subspace of coincident
two vortices in $U(N)$ gauge theory is therefore  found to be a
weighted Grassmannian  manifold, \beq
 {\cal M}_{N,k=2}|_{\rm coincident}
  \simeq W Gr_{{N+1,2}}^{({\small  {  1,\cdots,1}}  ,  0) }.
  \label{eq:w-Grassmann}
\eeqq
We note again that the elements
${\cal S} = - {\bf 1}_2$ and $\lam = -1$
acting as
\beq
 \left(\begin{array}{cc}
  {\bf \Psi}^T & v
 \end{array}
 \right)
 \to
 \left(\begin{array}{cc}
  {\bf \Psi}^T & - v
 \end{array}
 \right)
\eeqq
is precisely the ${\bf Z}_2$ action
which gives orbifold singularities.   
Note that, although the ordinary complex Grassmannian manifold 
$Gr_{{N+1,2}} \simeq \frac{SU(N+1)}{SU(N-1)\times SU(2)\times U(1)}$ 
naturally enjoys an $SU(N+1)$ isometry, the weighted Grassmannian
manifold $WGr_{{N+1,2}}^{(1,\cdots,1,0)}$ can have an $SU(N)$ 
isometry at most, due to the difference of $U(1)^{\bf C}$ charges.
This is consistent with the existence of 
the $SU(N)_{{\rm G+F}}$ symmetry acting on the moduli space 
in the $U(N)$ case.  
In cases of $N>2$ the orbifold singularities  
are not isolated points but form a submanifold given by $v=0$, 
which is the ordinary complex Grassmannian 
manifold $Gr_{N,2}\subset WGr_{{N+1,2}}^{(1,\cdots,1,0)}$ 
reflecting the $SU(N)_{{\rm G+F}}$ symmetry.

In Appendix~\ref{matrix}, beside giving the general procedure to
pass from the moduli matrix $H_0(z)$ to the K\"ahler quotient
construction, we directly show a one-to-one correspondence among the
patches of the moduli matrix for the $U(N)$, $k=2$ vortices and
those of the weighted Grassmannian manifold $W
Gr_{{N+1,2}}^{({\small  { 1,\cdots,1}}  ,  0) }$ and verify that the
transition functions of the latter perfectly match the ones obtained
from $H_0(z)$. We enforce this way the above result based on general
grounds (specifically the equivalence of moduli matrix and K\"ahler
quotient approach discovered in \cite{Eto:2006pg}).

\subsection{The case of $U(2)$ revisited}

As an illustration consider again the case of $U(2)$ theory.
 $2$ by $2$ matrices ${\bf Z}$ and ${\bf \Psi}$
 correspond to the moduli space of $k=2$ non-Abelian vortices
in the $U(2)$
gauge theory with the equivalence relation Eq.~(\ref{eq:MSbyZPsi}).
The double co-axial vortices are described by $\det\left(z{\bf 1}_2 - {\bf Z}\right) = z^2$.
This can be rewritten as $\Tr {\bf Z} = \Tr {\bf Z}^2 = 0$.
These conditions are easily solved and   we find that
these  vortices are described by the following two 2 by 2 matrices
${\bf Z}$ and ${\bf \Psi}$:
\beq
{\bf Z} = \epsilon \, v \,  v^T =
\left(
\begin{array}{cc}
v_1 \, v_2 & v_2^2 \\
- v_1^2 & -v_1 \, v_2
\end{array}
\right),\quad
{\bf \Psi}
= \left(
\begin{array}{cc}
\psi_{11} & \psi_{12}\\
\psi_{21} & \psi_{22}
\end{array}
\right)
\label{eq:z_psi_k2N2}
\eeq
where $v^T = (v_1,\ v_2)$ and
$\epsilon = \left(\begin{smallmatrix} 0 & 1 \\ -1 & 0 \end{smallmatrix}\right)$.
These  obey the equivalence relation $GL(2,{\bf C})$ given in Eq.~(\ref{eq:VZPsi}).
At this stage the matter would become simple if we consider $v$ rather than ${\bf Z}$.
Since ${\bf Z}$ is in the adjoint representation of $GL(2,{\bf C})$, $v$ is in
the fundamental representation of $SL(2,{\bf C})$. Notice that $v$
is not charged under  the overall $U(1)^{\bf C} \subset GL(2,{\bf C})$.

It is natural to define $k \, (=2)$ by $N+1 \, (=2+1)$ matrix \beq M
= \left({\bf \Psi}^T,\ v \right) = \left(
\begin{array}{ccc}
\psi_{11} & \psi_{21} & v_1\\
\psi_{12} & \psi_{22} & v_2
\end{array}
\right).
\label{eq:M32}
\eeq
This matrix $M$ transforms under $GL(2,{\bf C}) = U(1)^{\bf C} \times SL(2,{\bf C})$
as follows
\beq
 M = \left(\Psi^T,\ v \right)
 \sim \left({\cal S} \, \lambda \, \Psi^T,\ {\cal S} \, v \right)
 \label{eq:wg32} 
\eeq 
where ${\cal S} \in SL(2,{\bf C})$ and $\lambda
\in U(1)^{\bf C}$. If the vector $v$ had a charge 1 under $U(1)^{\bf
C}$  (that is, $v \to \lambda \, v$), the above identification would
correspond to the complex Grassmannian $Gr_{3,2} \simeq M/GL(2,{\bf
C})$ which is same as ${\bf C}P^2$. But since  $v$ is not charged
under $U(1)^{\bf C}$,    the manifold is not a Grassmannian.
 Eq.~(\ref{eq:wg32}) is an example of a
weighted Grassmannian   manifold  and we denote  it by
$WGr_{3,2}^{(1,1,0)}$. Here the numbers $(1,1,0)$ denote the
$U(1)^{\bf C}$-charges (the weights) of columns of $M$.

We choose appropriate  $GL(2,{\bf C})$ matrices  to obtain various
patches on the moduli space for the composite  vortices. Let us
define the 2 by 2 minors $M_{[ij]}$ and their determinants as \beq
M_{[ij]} = \left(
\begin{array}{cc}
M_{i1} & M_{j1} \\
M_{i2} & M_{j2}
\end{array}
\right), \quad \tau _{ij} =\det M_{[ij]} . \label{eq:def_minor} \eeq
There are  3 minors $M_{[12]}$, $M_{[23]}$ and $M_{[13]}$. Using the
$GL(2,{\bf C})$, one of them can be brought to
 identity. So  one has  3 patches as follows.
\begin{itemize}
\item $M_{[23]} = {\bf 1}_2$ patch.\\
First act
${\cal S} = \left(
\begin{smallmatrix} \tau_{23} & 0 \\ 0 & 1 \end{smallmatrix} \right) M_{[23]}^{-1}$
to the matrix $M$ in Eq.~(\ref{eq:M32}), and after that
 by  $\lambda = \tau_{23}^{-1}$:
\beq
M \to
\left(
\begin{array}{ccc}
\tau_{13} & \tau_{23} & 0\\
- \frac{\tau_{12}}{\tau_{23}}& 0 & 1
\end{array}
\right)
\to
\left(
\begin{array}{ccc}
\frac{\tau_{13}}{\tau_{23}} & 1 & 0 \\
- \frac{\tau_{12}}{\tau_{23}^2} & 0 & 1
\end{array}
\right).
\eeq
\item $M_{[12]} = {\bf 1}_2$ patch.\\
First one  acts ${\cal S} = \left(\tau_{12}\right)^{1/2}
M_{[12]}^{-1}$ to the matrix $M$ in Eq.~(\ref{eq:M32}), and after
that then by  $\lambda = \left(\tau_{12}\right)^{-\frac{1}{2}}$:
\beq M \to \left(
\begin{array}{ccc}
\sqrt{\tau_{12}} & 0 &
\frac{- \tau_{23}}{\sqrt{\tau_{12}}}\\
0 & \sqrt{\tau_{12}} & \frac{\tau_{13}}{\sqrt{\tau_{12}}}
\end{array}
\right)
\to
\left(
\begin{array}{ccc}
1 & 0 & \frac{- \tau_{23}}{\sqrt{\tau_{12}}}\\
0 & 1 &   \frac{\tau_{13}}{\sqrt{\tau_{12}}}
\end{array}
\right).
\eeq
Notice that the  element ${\cal S}=-{\bf 1}_2$ with $\lambda = -1$, has not been fixed,
thus  one has a ${\bf Z}_2$ symmetry in this patch
\beq
\left(
\begin{array}{c}
- \frac{ \tau_{23}}{\sqrt{\tau_{12}}}\\
\frac{\tau_{13}}{\sqrt{\tau_{12}}}
\end{array}
\right) \sim
\left(
\begin{array}{c}
\frac{\tau_{23}}{\sqrt{\tau_{12}}}\\
- \frac{\tau_{13}}{\sqrt{\tau_{12}}}
\end{array}
\right)
\eeq
\item $M_{[13]} = {\bf 1}_2$ patch.\\
Act  first by
${\cal S} = \left(
\begin{smallmatrix} \tau_{13} & 0 \\ 0 & 1 \end{smallmatrix} \right) M_{[13]}^{-1}$
to the matrix $M$ in Eq.~(\ref{eq:M32}), and then by
  $\lambda = \tau_{13}^{-1}$:
\beq
M \to
\left(
\begin{array}{ccc}
\tau_{13} & \tau_{23} & 0\\
0 &  \frac{\tau_{12}}{\tau_{13}} & 1
\end{array}
\right)
\to
\left(
\begin{array}{ccc}
1 & \frac{\tau_{23}}{\tau_{13}} &  0 \\
0 & \frac{\tau_{12}}{\tau_{13}^2} & 1
\end{array}
\right)
\eeq
\end{itemize}
The corresponding matrices $\left(\begin{smallmatrix}{\bf
\Psi}\\{\bf Z}\end{smallmatrix}\right)$ for the above three patches
are summarized as follows 
\beq
\left(
\begin{array}{cc}
\frac{\tau_{13}}{\tau_{23}} & \frac{- \tau_{12}}{\tau_{23}^2}  \\
1 & 0 \\
0 & 1 \\
0 & 0
\end{array}
\right),
\left(
\begin{array}{cc}
1 & 0\\
0 & 1\\
\frac{- \tau_{23} \tau_{13}}{\tau_{12}} & \frac{\tau_{13}^2}{\tau_{12}} \\
\frac{- \tau_{23}^2}{\tau_{12}} & \frac{\tau_{23}
\tau_{13}}{\tau_{12}}
\end{array}
\right),
\left(
\begin{array}{ccc}
1 & 0 \\
\frac{\tau_{23}}{\tau_{13}} & \frac{\tau_{12}}{\tau_{13}^2} \\
0 & 1 \\
0 & 0
\end{array}
\right).
\label{eq:ZP_patch}
\eeq
The leftmost  one corresponds to the $M_{[23]}={\bf 1}_2$ patch,
the middle to the $M_{[12]}={\bf 1}_2$    and
the rightmost one  to the $M_{[13]}={\bf 1}_2$ patch.
Clearly, these should be identified with the matrices in Eq.~(\ref{eq:HofN2k2-3})
which were obtained from the moduli matrices  $H_0^{(2,0)}(z)$,
$H_0^{(1,1)}(z)$ and $H_0^{(0,2)}(z)$ given in Eq.~(\ref{eq:HofN2k2})
through the relation Eq.~(\ref{eq:Z_Psi}).
Therefore,
$M_{[23]}={\bf 1}_2$ patch and $\tilde {\cal U}^{(0,2)}$ patch,
$M_{[12]}={\bf 1}_2$ patch and $\tilde {\cal U}^{(1,1)}$ patch, and
$M_{[13]}={\bf 1}_2$ patch and $\tilde {\cal U}^{(2,0)}$ patch.
The concrete identification is
\beq
\left(
\begin{array}{c}
\frac{\tau_{12}}{\tau_{13}^2}\\
\frac{\tau_{23}}{\tau_{13}}
\end{array}
\right)
=
\left(
\begin{array}{c}
-a\\
b
\end{array}
\right),\
\left(
\begin{array}{c}
\frac{\tau_{13}}{\sqrt{\tau_{12}}}\\
\frac{\tau_{23}}{\sqrt{\tau_{12}}}
\end{array}
\right)
=
\left(
\begin{array}{c}
X\\
Y
\end{array}
\right),\
\left(
\begin{array}{c}
\frac{\tau_{12}}{\tau_{23}^2}\\
\frac{\tau_{13}}{\tau_{23}}
\end{array}
\right)
=
\left(
\begin{array}{c}
a'\\
b'
\end{array}
\right).
\eeq

Now we are ready to understand a little mysterious relation
Eq.~(\ref{eq:equiv}) which gave us the weighted complex projective
space $W{\bf C}P^2_{(2,1,1)}$. Ordinary complex Grassmannian
$Gr_{3,2}$ is known to be equivalent to ${\bf C}P^2$ and the
weighted cases are quite analogous. Because  all the parameters in
Eq.~(\ref{eq:ZP_patch}) are functions of the determinants of the
minors $M_{[12]}$, $M_{[23]}$ and $M_{[13]}$, it would be natural to
consider that the manifold is naturally parameterized by them. In
particular  the origin  of the  weighted equivalence relation
Eq.~(\ref{eq:equiv}) becomes clear  since $\tau_{ij}$ are invariant
under $SL(2,{\bf C})$ while transforming  under $U(1)^{\bf C}$ as
\beq \left(
\begin{array}{c}
\phi_0\\
\phi_1\\
\phi_2
\end{array}
\right)
=
\left(
\begin{array}{c}
\tau_{12}\\
\tau_{13}\\
\tau_{23}
\end{array}
\right)
\sim
\left(
\begin{array}{c}
\lambda^2 \,\tau_{12}\\
\lambda \, \tau_{13}\\
\lambda \,  \tau_{23}
\end{array}
\right). \eeq This is nothing but  Eq.~(\ref{eq:equiv}). The $\tilde
{\cal U}^{(0,2)}$ patch is obtained by fixing with $\lambda =
\tau_{12}^{-1/2}$, the $\tilde {\cal U}^{(1,1)}$ patch  by $\lambda
= \tau_{23}^{-1/2}$ and the $\tilde {\cal U}^{(2,0)}$ patch by
$\lambda = \tau_{13}^{-1/2}$. Thus \beq \tilde {\cal M}_{N=2,k=2}
\simeq
 W{\bf C}P^2_{(2,1,1)} \simeq WGr_{3,2}^{(1,1,0)}.
\eeq

\subsection{The case of $U(3)$}

The moduli space of the co-axial  $k=2$  vortices in the $U(3)$
gauge theory is the weighted complex Grassmannian $\tilde {\cal
M}_{N=3,k=2} \simeq WGr_{4,2}^{(1,1,1,0)}$ as shown  already for general
$N$. The weighted Grassmannian is covered by $_4C_2 = 6$ patches as
the ordinary Grassmannian $Gr_{4,2}$. The patches are obtained as
follows. Let us begin with 2 by 4 matrix $M$ \beq M = \left(\Psi^T,\
v\right) = \left(
\begin{array}{cccc}
\psi_{11} & \psi_{21} & \psi_{31} & v_1 \\
\psi_{12} & \psi_{22} & \psi_{32} & v_2
\end{array}
\right) \eeq with an $GL(2,{\bf C}) = SL(2,{\bf C}) \times U(1)^{\bf
C}$ weighted equivalence relation \beq M = \left(\Psi^T,\ v\right)
\sim \left({\cal S} \, \lambda\ \, \Psi^T,\ {\cal S} \, v\right)
\eeq with ${\cal S}\in SL(2,{\bf C})$ and $\lambda \in U(1)^{\bf
C}$. The matrix $M$ has 6 minor matrices $M_{[ij]}$ whose size are 2
by 2 as given in Eq.~(\ref{eq:def_minor}). By using the $GL(2,{\bf
C})$, we can bring one of the 6 minors to the unit matrix.
\begin{itemize}
\item $M_{[a4]} = {\bf 1}_2$ patches ($a=1,2,3$):\\
In order to obtain $M_{[a4]} = {\bf 1}_2$ patch, one must  perform
${\cal S}_a =
\left(
\begin{smallmatrix} \tau_{a4} & 0 \\ 1 & 0 \end{smallmatrix}
\right) M_{[a4]}^{-1} \in SL(2,{\bf C})$ and $\lambda_a =
\tau_{a4}^{-1} \in U(1)^{\bf C}$. For example, $M_{[14]} = {\bf
1}_2$ patch is obtained as follows: \beq M \to \left(
\begin{array}{cccc}
\tau_{14} & \tau_{24} & \tau_{34} & 0 \\
0 & \frac{\tau_{12}}{\tau_{14}} & \frac{\tau_{13}}{\tau_{14}} & 1
\end{array}
\right)\to
\left(
\begin{array}{cccc}
1 & \frac{\tau_{24}}{\tau_{14}} & \frac{\tau_{34}}{\tau_{14}} & 0 \\
0 & \frac{\tau_{12}}{\tau_{14}^2} & \frac{\tau_{13}}{\tau_{14}^2} &
1
\end{array}
\right).
\eeq
Similarly one finds  the remaining $M_{[24]} = {\bf 1}_2$
and $M_{[34]} = {\bf 1}_2$ patches
\beq
\left(
\begin{array}{cccc}
\frac{\tau_{14}}{\tau_{24}} & 1 & \frac{\tau_{34}}{\tau_{24}} & 0 \\
-\frac{\tau_{12}}{\tau_{24}^2} & 0 & \frac{\tau_{23}}{\tau_{24}^2} &
1
\end{array}
\right),\quad
\left(
\begin{array}{cccc}
\frac{\tau_{14}}{\tau_{34}} & \frac{\tau_{24}}{\tau_{34}} & 1 & 0 \\
- \frac{\tau_{13}}{\tau_{34}^2} & - \frac{\tau_{23}}{\tau_{34}^2} &
0 & 1
\end{array}
\right),
\eeq
respectively.
\item $M_{[ab]} = {\bf 1}_2$ patches ($a,b = 1,2,3,$ $\, a<b$):
In order to get  a $M_{[ab]} = {\bf 1}_2$ patch, one must make a
transformation with ${\cal S}_{ab} = \left( \tau_{ab}
\right)^{\frac{1}{2}} M_{[ab]}^{-1} \in SL(2,{\bf C})$ and
$\lambda_{ab} = \left(\tau_{ab}\right)^{-\frac{1}{2}} \in U(1)^{\bf
C}$. For example, $M_{[12]} = {\bf 1}_2$ patch is obtained as
follows: \beq \hspace*{-1cm} M \to \left(
\begin{array}{cccc}
\sqrt{\tau_{12}} & 0 & \frac{-\tau_{23}}{\sqrt{\tau_{12}}}
& \frac{-\tau_{24}}{\sqrt{\tau_{12}}} \\
0 & \sqrt{\tau_{12}} & \frac{\tau_{13}}{\sqrt{\tau_{12}}} &
\frac{\tau_{14}}{\sqrt{\tau_{12}}}
\end{array}
\right)
\to
\left(
\begin{array}{cccc}
1 & 0 & \frac{-\tau_{23}}{\tau_{12}} &
\frac{-\tau_{24}}{\sqrt{\tau_{12}}} \\
0 & 1 & \frac{\tau_{13}}{\tau_{12}} & \frac{\tau_{14}}
{\sqrt{\tau_{12}}}
\end{array}
\right).
\eeq
The patches $M_{[13]} = {\bf 1}_2$
and $M_{[23]} = {\bf 1}_2$ can be obtained  similarly
\beq
\left(
\begin{array}{cccc}
1 & \frac{\tau_{23}}{\tau_{13}}  & 0
& -\frac{\tau_{34}}{\sqrt{\tau_{13}}} \\
0 & \frac{\tau_{12}}{\tau_{13}} & 1 &
\frac{\tau_{14}}{\sqrt{\tau_{13}}}
\end{array}
\right),\quad
\left(
\begin{array}{cccc}
\frac{\tau_{13}}{\tau_{23}} &  1 & 0
& -\frac{\tau_{34}}{\sqrt{\tau_{23}}} \\
-\frac{\tau_{12}}{\tau_{23}} & 0 & 1 &
\frac{\tau_{24}}{\sqrt{\tau_{23}}}
\end{array}
\right).
\eeq
Notice that  ${\cal S}=-{\bf 1}_2,\ \lambda = -1$  have not used up, so
one has a  ${\bf Z}_2$ symmetry in these patches.
\end{itemize}

$\tau_{ab}$ ($a,b=1,2,3,4$ with $a<b$) can be seen as natural
coordinates for the manifold $WGr_{4,2}^{(1,1,1,0)}$.
 $\lambda {\cal S} \in GL(2,{\bf C})$ equivalence relation is expressed on these
coordinates as
\beq
\left(
\begin{array}{c}
\tau_{12}\\
\tau_{23}\\
\tau_{13}\\
\tau_{14}\\
\tau_{24}\\
\tau_{34}
\end{array}
\right)
\sim
\left(
\begin{array}{c}
\lambda^2 \,\tau_{12}\\
\lambda^2 \,\tau_{23}\\
\lambda^2 \, \tau_{13}\\
\lambda \, \tau_{14}\\
\lambda \,  \tau_{24}\\
\lambda  \, \tau_{34}
\end{array}
\right). \label{eq:equiv42} \eeq 
This weighted equivalence relation
for 6 complex parameters defines the weighted complex projective
space $W{\bf C}P^5_{(2,2,2,1,1,1)}$ whose complex dimension is 5.
The $W{\bf C}P^5_{(2,2,2,1,1,1)}$ is a bigger manifold than the
$WGr_{4,2}^{(1,1,1,0)}$ whose complex dimension is 4. The embedding
relation $WGr_{4,2}^{(1,1,1,0)} \subset W{\bf C}P^5_{(2,2,2,1,1,1)}$
is the same as that for the ordinary Grassmannian to the ordinary
complex projective space through the so-called  Pl\"ucker relation
\beq
\begin{array}{c}
\tau_{12}\tau_{34} - \tau_{13}\tau_{24} + \tau_{14}\tau_{23} = 0.
\end{array} \label{eq:Plucker}
\eeq
The ${\bf Z}_2$ symmetry on the $W{\bf C}P^5_{(2,2,2,1,1,1)}$ is now realized as
the action of $\lambda = -1$ which acts as
\beq
\left(
\begin{array}{c}
\tau_{12}\\
\tau_{23}\\
\tau_{13}\\
\tau_{14}\\
\tau_{24}\\
\tau_{34}
\end{array}
\right)
\sim
\left(
\begin{array}{c}
\tau_{12}\\
\tau_{23}\\
\tau_{13}\\
-\tau_{14}\\
-\tau_{24}\\
-\tau_{34}
\end{array}
\right). \eeq The patches $M_{[a4]} = {\bf 1}_2$  ($a=1,2,3$) can be
obtained  fixing the $U(1)^{\bf C}$ ambiguity by choosing $\lambda =
\left(\tau_{a4}\right)^{-1}$ while the patches $M_{[ab]} = {\bf
1}_2$ ($a,b=1,2,3$ with $a<b$)  are obtained by choosing $\lambda =
\left(\tau_{ab}\right)^{-\frac{1}{2}}$.
Note that one can easily confirm that 
the whole of the orbifold singularities 
of the ${\bf Z}_2$ action form a submanifold 
${\bf C}P^2 \simeq Gr_{3,2}\subset WGr_{4,2}^{(1,1,1,0)}$, 
since the condition $\tau_{14}=\tau_{24}=\tau_{34}=0$ for 
the singularity solves 
the Pl\"ucker condition Eq.~(\ref{eq:Plucker}) and 
the unconstrained parameters $(\tau_{12},\tau_{23},\tau_{13})$ 
have the ordinary equivalence relation 
as that of ${\bf C}P^2$ with $\lambda'=\lambda^2$ 
in Eq.~(\ref{eq:equiv42}).

For completeness we list the moduli matrices in the case of $k=2$ and $N=3$.
The  six patches
$\tilde{\cal U}^{(2,0,0)}$, $\tilde{\cal U}^{(0,2,0)}$,
$\tilde{\cal U}^{(0,0,2)}$, $\tilde{\cal U}^{(1,1,0)}$,
$\tilde{\cal U}^{(0,1,1)}$ and $\tilde{\cal U}^{(1,0,1)}$
of the  $k=2$ co-axial vortices in the $U(3)$ gauge theory are
  given by
\begin{alignat}{3}
&H_0^{(0,0,2)}(z) =
\left(
\begin{array}{ccc}
1 & 0 & - a_1 z - b_1 \\
0 & 1 & - a_2 \, z - b_2 \\
0 & 0 & z^2
\end{array}
\right),
&\quad
&H_0^{(1,1,0)}(z) =
\left(
\begin{array}{ccc}
z + X \, Y & - {X}^2 & 0 \\
{Y}^2 & z - X \, Y & 0 \\
-\gamma & -\chi & 1
\end{array}
\right),\\
&H_0^{(0,2,0)}(z) =
\left(
\begin{array}{ccc}
1 & - a'_1 \, z - b'_1 & 0 \\
0 & z^2 & 0 \\
0 & - a'_2 \,z - b'_2 & 1
\end{array}
\right),
&\quad
&H_0^{(1,0,1)}(z) =
\left(
\begin{array}{ccc}
z + X' \, Y' & 0 & - {X'}^2 \\
- \gamma' & 1 &  -\chi'\\
{Y'}^2 & 0 & z - X' \, Y'
\end{array}
\right),\\
&H_0^{(2,0,0)}(z) =
\left(
\begin{array}{ccc}
z^2 & 0 & 0 \\
- a''_1 \, z - b''_1 & 1 & 0  \\
- a''_2 \, z - b''_2 & 0 & 1
\end{array}
\right),
&\quad
&H_0^{(0,1,1)}(z) =
\left(
\begin{array}{ccc}
1 & - \gamma'' & - \chi'' \\
0 & z + X'' \, Y'' & - {X''}^2 \\
0 & {Y''}^2 & z - X'' \,Y''
\end{array}
\right),
\end{alignat}
with   $\left(X,Y\right)\sim \left(-X,-Y\right)$,
$\left(X',Y'\right)\sim \left(-X',-Y'\right)$ and
$\left(X'',Y''\right)\sim \left(-X'',-Y''\right)$. 
These  identifications  lead to the orbifold singularities 
along ${\bf C}P^2$, as we mentioned,  which is parameterized by 
three patches $(\gamma,\chi), (\gamma',\chi')$ 
and $(\gamma'',\chi'')$. 
The determinant
of each of these matrices is equal to $z^2$ corresponding to the
fact that these describe double  vortices one sitting on the other,
at the origin of the $z$ plane. The transition functions and other
details are given in Appendix~\ref{matrix}.

\section{Conclusion}

In this paper we have studied and determined the structure 
of the moduli space of  certain composite
 non-Abelian vortices, appearing in $U(N)$ gauge theories 
in the Higgs phase.
The moduli subspace of  two  co-axial vortices (or equivalently,
axially symmetric $k=2$  vortices) in  the $U(N)$ gauge theories
with $N$ flavors, is  found to be a weighted Grassmannian manifold,
Eq.~(\ref{eq:w-Grassmann}). In the case of $U(2)$ gauge theory, it
reduces to a weighted projective space 
$W{\bf C}P^2_{(2,1,1)} \simeq {\bf C}P^{2}/{\bf Z}_2$ 
(${\bf C}P^2$ homotopically), 
in agreement with the known results
\cite{Hashimoto:2005hi, Auzzi:2005gr}. 
This space contains a ${\bf Z}_2$ orbifold
(conic) singularity at the origin of the $(1,1)$ patch.
In the case of $U(N)$ gauge theory, 
it contains singularities along $Gr_{N,2}$. 

The presence of this kind of orbifold singularities is a general
feature of weighted Grassmannian manifold. This fact implies the
necessity to reconsider the reconnection of non-Abelian vortices. So
far this issue has been studied considering the moduli space of
$k=2$ co-axial vortices smooth everywhere \cite{Hashimoto:2005hi}.
We claim that this is not the case and that we need to analyze the
metric on the $k=2$ vortices moduli space to address the problem.

An interesting question is how our results are generalized in the
case of semi-local non-Abelian vortices
\cite{Shifman:2006kd,Eto:2006pg}. Extension of the results of this
paper to the semi-local cases will be discussed elsewhere.

It  would be interesting also to extend our study to vortices of
different kind, such as those appearing in  $SO(N)$ theories
\cite{Ferretti:2006jg}.

The  implications of our results on the properties of non-Abelian
{\it monopoles}  appearing in related systems, will be discussed in
a separate paper.

\section*{Acknowledgments}
We would like to thank Koji Hashimoto for useful discussions. 
One of the authors (K.~K.) thanks H. Kawai and M. Ninomiya for a kind
invitation to RIKEN and Yukawa Institute for Theoretical Physics
(April 2006) where some of the work has been done, and Roberto
Auzzi, Giampiero Paffuti,  Pietro Menotti for discussions.
M.~E., M.~N. and K.~O. wish to thank the theoretical high 
energy physics group of University of Pisa  and INFN, Sezione di 
Pisa, for their hospitality, 
where this work was 
initiated. 
The work of M.~E. and K.~O. is supported by Japan 
Society for the Promotion of Science under the Post-doctoral Research Program. 
The work of N.~Y. is supported by the Special Postdoctoral Researchers 
Program at RIKEN.

\appendix
\section{Relation to ASY ansatz} \label{ASY}
The ansatz in \cite{Auzzi:2005gr}   is formulated   
in terms of  two independent unit
vectors, $\vec{n}_{1}$ and $\vec{n}_{2}$. Using a global
color-flavor rotation  the two vectors can be rotated  into the
following form:
\begin{equation}\label{ASY vector}
\vec{n}_{1}=(0,0,1), \quad \vec{n}_{2}=(\sin \alpha,0,\cos \alpha)
\end{equation}
where $\alpha$ is the relative angle between $\vec{n}_{1}$ and
$\vec{n}_{2}$. Now it is straightforward to derive the moduli matrix
$H_{0}$ that corresponds to this particular choice of parameters, as
was done in \cite{Auzzi:2005gr}:
\begin{equation}\label{ASY ansatz}
      H_{0}(z,\alpha)= \left( \begin{array}{cc}
       -\cos \frac{\alpha}{2} \, z^2 & \sin \frac{\alpha}{2} \, z \\
        -\sin \frac{\alpha}{2} \, z & -\cos \frac{\alpha}{2}
     \end{array} \right).
\end{equation}
This matrix can be put  into an upper-right  triangular form
\begin{equation}
      H_{0}(z,\alpha)= \left( \begin{array}{cc}
       z & \cot \frac{\alpha}{2} \\
        0 & z
     \end{array} \right)\label{eq:ASYtriang}, \end{equation}
     by a $V$ transformation.

The ASY vortices with generic orientation vectors $\vec{n}_{1}$ and
$\vec{n}_{2}$ can be found simply by an overall $SO(3)$  rotation of
the above.  To find the moduli matrix representation of the general
ASY ansatz,  we must go to the system in which
\begin{equation}\label{n1}
    \vec{n}_{1}=(- \sin \alpha_1 \, \cos \beta_1, \, \sin \alpha_1 \, \sin \beta_1, \,\cos \alpha_1).
\end{equation}
To obtain such  a general ASY solution, parameterized with four
angular coordinates:
\begin{equation}\label{n1'}
    H_{0}(z,\alpha, \beta,\alpha_1, \beta_1)
\end{equation}
where $(\alpha_1, \beta_1)$ represent the orientation ${\bf n}_{1}$
while the angles $(\alpha, \beta)$ stand for  the orientation of the
vector ${\bf n}_{2}$ relative  to ${\bf n}_{1}$, we rotate the
moduli matrix Eq.~(\ref{eq:ASYtriang}) with a global rotation matrix:
\begin{eqnarray}
U &=&\exp( \frac{i}{2}  \eta_1 \tau_3) \, \exp(\frac{i}{2} \alpha_1
\tau_2) \, \exp( \frac{i}{2} \beta_1 \tau_3), \nonumber   \\
\label{eq:rot}
   &=&   \left(\begin{array}{cc}
             e^{ \frac{i}{2}  (\beta_1+\eta_1)}\cos \frac{\alpha_1}{2} & -e^{-\frac{i}{2}  (\beta_1-\eta_1)}\sin \frac{\alpha_1}{2} \\
             e^{\frac{i}{2}  (\beta_1-\eta_1)}\sin \frac{\alpha_1}{2}& e^{-\frac{i}{2}  (\beta_1+\eta_1)}\cos \frac{\alpha_1}{2}
          \end{array}
    \right),
\end{eqnarray}
where $(\eta_1,\alpha_1, \beta_1)$ are the Euler angles. After the
rotation
\begin{equation}
    H_0(z,\alpha, \eta_1,\alpha_1, \beta_1) = H_0(z,\alpha) \, U(\eta_1,\alpha_1,
    \beta_1),
\end{equation}
we put the result into the upper-right triangular form, $H_0(z,\alpha,
\eta_1,\alpha_1, \beta_1) \sim V\, H_0(z,\alpha, \eta_1,\alpha_1,
\beta_1)$, to get
\begin{equation} H_0(z,\alpha, \eta_1,\alpha_1,
\beta_1)=\label{asyident}
    \left( \begin{array}{cc}
             1 & -e^{- i \beta_1}\cot \frac{\alpha_1}{2} -z e^{- i (\beta_1- \eta_1)}\csc^2 \frac{\alpha_1}{2}\, \tan \frac{\alpha}{2}\\
             0 & z^2
           \end{array}
    \right).
\end{equation}
The $V$ transformation needed is
\begin{equation}\label{vtrans}
    V=\left(\begin{array}{cc}
            e^{- i/2 (\beta_1-\eta_1)}\csc \frac{\alpha_1}{2} \, \tan \frac{\alpha}{2} & -e^{- i/2 (\beta_1-3 \eta_1)}\cot \frac{\alpha_1}{2} \, \csc \frac{\alpha_1}{2} \, \tan \frac{\alpha}{2} \\
            -z \,e^{- i/2 (\beta_1-\eta_1)}\sin \frac{\alpha_1}{2} \, \tan \frac{\alpha}{2} & z \, e^{ i/2 \beta_1} \cos
            \frac{\alpha_1}{2}+e^{ i/2 (\beta_1-\eta_1)}  \sin
            \frac{\alpha_1}{2}
          \end{array}
     \right).
\end{equation}
With an arbitrary choice for the origin of the $\beta$ angle we can
identify   $\beta=\eta_1$.  Thus the ASY ansatz has the moduli
matrix representation,  with
\begin{equation}\label{final}
    b=e^{- i \beta_1}\cot \frac{\alpha_1}{2},\quad a= e^{- i (\beta_1- \beta)}\csc^2 \frac{\alpha_1}{2} \, \tan \frac{\alpha}{2}
\end{equation}
Note that  $a \to \infty $  as $\alpha \to \pi$ and we get the
singlet point of the moduli space; while $a \to 0$  as $\alpha \to 0
$ and we get a ``doublet'' transforming as a $k=1$ vortex with the
orientation vector ${\bf n}_{1}$, in accord with ASY and with our
results.

\section{Relation to HT-ASY analysis} \label{HTASY}
The K\"ahler quotient construction 
(D-brane construction by Hanany-Tong) for the
moduli space of $k$ vortices in $U(N)$ gauge theory coupled with $N$
Higgs fields is given by the two matrices $Z$ and $\psi$. Here $Z$
is $k$ by $k$ matrix and $\psi$ is $N$ by $k$ matrix which satisfy
the following constraint
\beq \left[ Z^\dagger, Z \right] +
\psi^\dagger\psi = {\bf 1}_2, \label{eq:d}
\eeq
and are divided by the $U(k)$ symmetry
\beq Z \to \tilde {\cal V} \,
Z \, \tilde {\cal V}^\dagger,\quad \psi \to  \psi \, \tilde {\cal
V}^\dagger,
\eeq
with $\tilde {\cal V} \in U(k)$. The eigenvalues of the matrix $Z$
are thought of as the positions of the vortices. After performing an
appropriate transformation $U(k)$, we can always bring the matrix
$Z$ into a triangle matrix. As a concrete example, let us consider
composing two vortices at the origin in the $U(2)$ model. The matrices
are of the form in that gauge: \beq Z = \left(
\begin{array}{cc}
0 & w \\
0 & 0
\end{array}
\right),\quad
\psi = \left(
\begin{array}{cc}
\vec A & \vec B
\end{array}
\right), \label{eq:asy} \eeq with $w\in {\bf C}$ and two complex
vectors $\vec A^T = \left(A_1, A_2\right)$ and $\vec B^T =
\left(B_1, B_2\right)$. Note that we have not fixed two $U(1)$
symmetries $e^{i\theta_1} \in U(1)_1$ and $e^{i\theta_2} \in U(1)_2$
which do not change the triangle form of the matrix $Z$: \beq \tilde
{\cal V} = \left(
\begin{array}{cc}
e^{i(\theta_1+\theta_2)} & 0\\
0 & e^{i(\theta_1 - \theta_2)}
\end{array}
\right).
\eeq
The charges of $\vec A$, $\vec B$ and $w$ under those $U(1)$'s are
summarized in the following table.
\begin{table}[ht]
\begin{center}
\ \\
\vspace*{-.2cm}
\begin{tabular}{c||c|c}
 & $U(1)_1$ & $U(1)_2$ \\
\hline
$\vec A$ & $-1$ & $-1$ \\
$\vec B$ & $-1$ & $1$ \\
$w$ & $0$ & $2$
\end{tabular}
\caption{\small The charges under $U(1)_1$ and $U(1)_2$.}
\end{center}
\end{table}
We have to fix these $U(1)$'s to realize the moduli space of the
composing two vortices. To this end, we first absorb the phase of
$w$ as $w = |w|$ by use of $U(1)_2$. Then we have only $U(1)_1$ as
unfixed gauge symmetry. Plugging Eq.~(\ref{eq:asy}) into
Eq.~(\ref{eq:d}), we obtain three constraints
\beq |\vec A|^2 -
|w|^2 = 1,\quad |\vec B|^2 + |w|^2 = 1,\quad \vec A ^\dagger \cdot
\vec B = 0. \label{eq:d_const}
\eeq
From the middle constraint of Eq.~(\ref{eq:d_const}), the vector
$\vec B$ can be written as the following form
\beq \vec B = \sqrt{1
- |w|^2} \left(
\begin{array}{c}
-e^{-i\beta_2}\sin\xi \\
e^{-i\beta_1}\cos\xi
\end{array}
\right).
\eeq
Then from the first and the last constraints in Eq.~(\ref{eq:d_const}),
the remaining vector $\vec A$ can be expressed as the following form
\beq
\vec A = \sqrt{1 + |w|^2} \, e^{i\alpha}\,  \epsilon \, \frac{\vec B
^*}{|\vec B|} = \sqrt{1 + |w|^2} \, e^{i\alpha} \left(
\begin{array}{c}
e^{i\beta_1}\cos\xi\\
e^{i\beta_2}\sin\xi
\end{array}
\right)
\eeq
with the antisymmetric tensor $\epsilon$ ($\epsilon^{12} = 1$).
Let us fix the remaining $U(1)_1$ gauge symmetry by choosing
$\theta_1 = \frac{\alpha}{2}$.
Finally, we get the following form
\beq
Z =
\left(
\begin{array}{cc}
0 & |w| \\
0 & 0
\end{array}
\right),\quad
\psi = \left(
\begin{array}{cc}
\sqrt{1+|w|^2} \, e^{i\gamma_1} \, \cos\xi & - \sqrt{1-|w|^2} \, e^{-i\gamma_2}\, \sin\xi\\
\sqrt{1+|w|^2} \, e^{i\gamma_2} \, \sin\xi & \sqrt{1-|w|^2} \,e^{-
i\gamma_1}\ \, \cos\xi
\end{array}
\right),
\label{eq:ht_asy}
\eeq
where $\gamma_1 = \frac{\alpha}{2}+\beta_1$
and $\gamma_2 = \frac{\alpha}{2}+\beta_2$.
At this stage, we have fixed both $U(1)_1$ and $U(1)_2$.
But we have to be careful because $U(1)_2$ has not been completely fixed  yet.
In fact, ${\bf Z}_2$ transformation by $\theta_2 = \pi$ is unfixed since $w$ has
charge 2 under $U(1)_2$ gauge symmetry ($w=|w| \to +|w|$).
 Therefore we need to take ${\bf Z}_2$ identification
into account
\beq
{\bf Z}_2:\ \ \psi \to - \psi.
\label{eq:z2_ht}
\eeq

Now we have completely fixed the $U(2)$ gauge symmetry and have reached
a patch of the K\"ahler quotient in which ${\bf Z}_2$ symmetry is equipped.
This patch should be identified with our matrices
${\bf Z}^{(1,1)}$ and ${\bf \Psi}^{(1,1)}$ in the $\tilde {\cal U}^{(1,1)}$ patch
given in Eq.~(\ref{eq:HofN2k2-3}) in which
the ${\bf Z}_2$ symmetry also exist.
Our matrices $\{{\bf Z}^{(1,1)},{\bf \Psi}^{(1,1)}\}$ and matrices $\{Z,\psi\}$ in
Eq.~(\ref{eq:ht_asy}) are transformed by $GL(2,{\bf C})$ transformation
\beq
{\cal V} = \psi \in GL(2,{\bf C}):\ \ {\bf Z}^{(1,1)} = \psi \, Z \,
\psi^{-1},\ {\bf \Psi}^{(1,1)} = \psi \, \psi^{-1} = {\bf 1}_2.
\eeq
More concretely, this can be written as
\beq
\left(
\begin{array}{cc}
-X \, Y & X^2 \\
-Y^2 & X \, Y
\end{array}
\right)
=
\frac{|w|\sqrt{1+|w|^2}}{\sqrt{1-|w|^2}}
\left(
\begin{array}{cc}
- e^{i(\gamma_1 + \gamma_2)}\sin\xi \, \cos\xi & e^{2i\gamma_1}\cos^2\xi\\
- e^{2i\gamma_1}\sin^2\xi & e^{i(\gamma_1 + \gamma_2)}\sin\xi \,
\cos\xi
\end{array}
\right).
\eeq
Thus we can find the relation
\beq
\left(
\begin{array}{c}
X\\
Y
\end{array}
\right)
= \sqrt{\frac{|w|\sqrt{1+|w|^2}}{\sqrt{1-|w|^2}}}
\left(
\begin{array}{c}
e^{i\gamma_1}\cos\xi \\
e^{i\gamma_2}\sin\xi
\end{array}
\right)
= \sqrt{\frac{|w|}{\sqrt{1-|w|^4}}} \ \vec A.
\eeq
Thus we conclude that our ${\bf Z}_2$ symmetry $(X,Y) \to -(X,Y)$ and the other
${\bf Z}_2$ symmetry ($\vec A \to - \vec A$)
in Eq.~(\ref{eq:z2_ht}) are completely equivalent.

To close completely the gap, here are the explicit relations  among
the HT parameters  ($w, \gamma_{1,2}, \xi$)   and those of ASY
construction  ($ \alpha, \beta, \alpha_{1}, \beta_{1}$) discussed in
the Appendix~\ref{ASY}: \beq \tan^{-1/2} \frac{\alpha}{2} =
\sqrt{\frac{|w|\sqrt{1+|w|^2}}{\sqrt{1-|w|^2}}}; \eeqq \beq
\gamma_{1} =  \frac{\beta_{1}- \beta  + \pi }{2}, \quad \gamma_{2} =
- \frac{\beta_{1} + \beta  - \pi }{2}, \quad
  \xi=\frac{\alpha_{1}}{2}.\eeqq

\section{Matrix representation} \label{matrix}

As was already emphasized, all the moduli parameters are contained
in the moduli matrix. The positions of $k$ vortices are given by the
zeros $\{z_i\}$ of the determinant $P(z) = \det H_0(z)$ of the
moduli matrix: $P(z=z_i) = 0$, and the orientations $\{\vec\phi_i\}$
are given by $H_0(z=z_i)\vec\phi_i = \vec 0$, a null vector at the
vortex positions in Eq.~(\ref{orientation}). This is a nice feature
of our approach, as long  as all the vortices are separated.
However, it does not give us a good  picture when the vortex axes
overlap,  as we have seen already.

In this Appendix  we will explain  a systematic method
to extract moduli parameters from the moduli matrix. A general
introduction to  this method was given in \cite{Eto:2006pg}.

\subsection{The case of $U(2)$}

Let us first extend the orientational vector $\vec \phi_i$ which is the {\it constant} vector
in Eq.~(\ref{orientation})
to a vector $\vec \phi_i(z)$ whose elements are not constants but
{\it holomorphic} polynomials of $z$   of order $ O(z^{k-1})$:
\beq
H_0(z) \,  \vec\phi_i(z)    =   J_{i}(z) \, P(z)  \equiv 0,    \quad Mod \,[P(z)].   \label{eq:phiz}
\eeq
for some holomorphic  $J_{i}(z)$.
This extended definition of the orientational vector reduces to Eq.~(\ref{orientation})
when we set $z=z_i$,   since $P(z=z_i)=0$.
The number of the linearly independent vectors $\vec\phi_i(z)$ is the same as the
degree of the polynomial $P(z)$, so that index $i$ runs from 1 to 2 for the $k=2$ vortices.

Introduce an $N$ by $k\,(=2)$ holomorphic matrix ${\bf \Phi}(z)$
from $\vec\phi_i(z)$ as \beq {\bf \Phi}(z) = \left(\vec\phi_1(z),\
\vec\phi_2(z)\right). \eeq Namely, ${\bf \Phi}(z)$ satisfies the
relation \beq H_0(z)\, {\bf \Phi}(z) \equiv 0,\quad 
{Mod}  \,\,[P(z)], \label{eq:defPhi} \qquad  P(z) = \det H_0(z). \eeq One  can
construct two constant  matrices ${\bf Z}$ which is a $k\,  (=2)$ by
$k \, (=2)$  matrix and ${\bf \Psi}$ which is a $N$ by $k \, (=2)$
matrix from ${\bf \Phi}(z)$ as follows. \beq z \, {\bf \Phi}(z) =
{\bf \Phi}(z)  \, {\bf Z} + {\bf \Psi} \, P(z). \label{eq:Z_Psi}
\eeq

For example, we can choose the following matrix satisfying
Eq.~(\ref{eq:defPhi}) with
the moduli matrix $H_0^{(0,2)}(z)$ in Eq.~(\ref{eq:HofN2k2})
\beq
{\bf \Phi}^{(0,2)}(z) =
\left(
\begin{array}{cc}
b\, z -b\, \alpha + a\, \beta & a\, z + b\\
z-\alpha & 1
\end{array}
\right). \eeq Here, $\vec \phi_2(z)^{\rm T}=(a\, z+b,1)$ is a
straightforward solution for Eq.~(\ref{eq:phiz}) since $\phi_2(z_i)$
$(i=1,2)$ are just the two orientational vectors given by
Eq.~(\ref{eq:vtx:fourkinds}) and $\vec \phi_1(z)$ is given by
$(z-\alpha)\, \vec \phi_2(z)$ with modulo $P(z) = z^2 - \alpha \, z
-\beta$. According to the prescription given in
Eq.~(\ref{eq:Z_Psi}), two matrices ${\bf Z}^{(0,2)}$ and
${\Psi}^{(0,2)}$ can be constructed  as follows:
\beq
z \, {\bf \Phi}^{(0,2)}(z) &=& \left(
\begin{array}{cc}
b\, z^2 -b\, \alpha z + a\, \beta z & a\, z^2 + b\, z\\
z^2-\alpha \, z & z
\end{array}
\right)
\nonumber\\
&=&
\left(
\begin{array}{cc}
b\, (z^2 - P(z)) -b\, \alpha \, z + a\, \beta\, z & a\, (z^2 - P(z)) + b\, z\\
(z^2 - P(z)) - \alpha \, z & z
\end{array}
\right)
+ P(z)
\left(
\begin{array}{cc}
b & a\\
1 & 0
\end{array}
\right)
\nonumber\\
&=&
\left(
\begin{array}{cc}
\beta (a \, z + b)& (a \, \alpha + b) z + a  \, \beta\\
\beta & z
\end{array}
\right)
+ P(z)
\left(
\begin{array}{cc}
b & a\\
1 & 0
\end{array}
\right)
\nonumber\\
&=&
{\bf \Phi}^{(0,2)}(z)
\left(
\begin{array}{cc}
0 & 1\\
\beta & \alpha
\end{array}
\right)
+ P(z)
\left(
\begin{array}{cc}
b & a\\
1 & 0
\end{array}
\right)
\label{eq:(0,2)}
\eeq
Thus we obtain ${\bf Z}^{(0,2)}$ and ${\Psi}^{(0,2)}$
corresponding to  $H_0^{(0,2)}(z)$:
\beq
{\bf Z}^{(0,2)} =
\left(
\begin{array}{cc}
0 & 1\\
\beta & \alpha
\end{array}
\right),\quad
{\bf \Psi}^{(0,2)} =
\left(
\begin{array}{cc}
b & a\\
1 & 0
\end{array}
\right).
\eeq
Let us turn our attention to another patch $H_0^{(1,1)}(z)$,  Eq.~(\ref{eq:HofN2k2}).
The corresponding orientational matrix ${\bf \Phi}^{(1,1)}(z)$  is
\beq
{\bf \Phi}^{(1,1)}(z)
= \left(
\begin{array}{cc}
z - \tilde\phi & \eta\\
\tilde\eta & z - \phi
\end{array}
\right). \eeq One can verify that $H_0^{(1,1)}(z) \, {\bf \Phi}(z)
\equiv {\bf 0}$ with modulo $P(z) = (z-\phi)(z-\tilde\phi) -\eta \,
\tilde\eta$. Again two matrices ${\bf Z}$ and ${\bf \Psi}$
satisfying Eq.~(\ref{eq:Z_Psi}) can be found: \beq z \,  {\bf
\Phi}^{(1,1)}(z) &=& \left(
\begin{array}{cc}
z^2 - \tilde\phi \, z & \eta \, z\\
\tilde\eta \, z & z^2 - \phi \, z
\end{array}
\right)\nonumber\\
&=&
\left(
\begin{array}{cc}
(z^2 - P(z)) - \tilde\phi\, z + P(z)& \eta \, z\\
\tilde\eta \, z & (z^2 - P(z)) - \phi \, z + P(z)
\end{array}
\right)\nonumber\\
&=&
\left(
\begin{array}{cc}
\phi \, z - \phi \,\tilde\phi + \eta\,\tilde\eta & \eta \, z\\
\tilde\eta \, z & \tilde \phi \, z - \phi\, \tilde\phi + \eta\,
\tilde\eta
\end{array}
\right)
+ P(z)
\left(
\begin{array}{cc}
1 & 0 \\
0 & 1
\end{array}
\right)
\nonumber\\
&=&
{\bf \Phi}^{(1,1)}(z)
\left(
\begin{array}{cc}
\phi & \eta \\
\tilde\eta & \tilde\phi
\end{array}
\right)
+  P(z)
\left(
\begin{array}{cc}
1 & 0 \\
0 & 1
\end{array}
\right)
\eeq
${\bf Z}^{(1,1)}$ and ${\bf \Psi}^{(1,1)}$,
\beq
{\bf Z}^{(1,1)} =
\left(
\begin{array}{cc}
\phi & \eta \\
\tilde\eta & \tilde\phi
\end{array}
\right),\quad
{\bf \Psi}^{(1,1)} =
\left(
\begin{array}{cc}
1 & 0 \\
0 & 1
\end{array}
\right)
\eeq
have the  same information as the moduli matrix $H_0^{(1,1)}(z)$.
Finally,
from  the  orientational matrix
\beq
{\bf \Phi}^{(2,0)}(z) =
\left(
\begin{array}{cc}
z - \alpha ' & 1 \\
b'\, z - b'\, \alpha' + a'\, \beta' & a'\, z + b'
\end{array}
\right).
\eeq
for the last patch $H_0^{(2,0)}(z)$ in Eq.~(\ref{eq:HofN2k2}),
one gets
\beq
{\bf Z}^{(2,0)} =
\left(
\begin{array}{cc}
0 & 1\\
\beta' & \alpha'
\end{array}
\right),\quad
{\bf \Psi}^{(2,0)} =
\left(
\begin{array}{cc}
1 & 0\\
b' & a'
\end{array}
\right).
\eeq

Summarizing,
\beq
\left(
\begin{array}{c}
{\bf \Psi}_{[N\times k]}\\
{\bf Z}_{[k \times k]}
\end{array}
\right)_{N=2,k=2}
=
\left(
\begin{array}{cc}
b & a\\
1 & 0\\
0 & 1\\
\beta & \alpha
\end{array}
\right),\
\left(
\begin{array}{cc}
1 & 0\\
0 & 1\\
\phi & \eta\\
\tilde\eta & \tilde\phi
\end{array}
\right),\
\left(
\begin{array}{cc}
1 & 0\\
b' & a'\\
0 & 1\\
\beta & \alpha
\end{array}
\right).
\eeqq
As was shown before, the moduli matrix $H_0^{(0,2)}(z)$,
$H_0^{(1,1)}(z)$ and $H_0^{(2,0)}(z)$ are connected by
$V$-equivalence relation ($H_0(z) \sim V(z) \, H_0(z)$ with $V(z)
\in GL(N=2,{\bf C})$). This leads to  the transition functions
between moduli parameters, see Eqs.(\ref{eq:02to11}) and
(\ref{eq:02to20}). From the view point of the matrices ${\bf Z}$ and
${\bf \Psi}$, the transition functions between them are given by the
$GL(k=2,{\bf C})$ equivalence relation
\beq
{\bf Z} \sim {\cal V} \, {\bf Z}\,  {\cal V}^{-1},\quad
{\bf \Psi} \sim {\bf \Psi} \, {\cal V}^{-1},
\label{eq:VZPsi}
\eeq
with ${\cal V} \in GL(k=2,{\bf C})$. This ${\cal V}$-equivalence
relation comes from ambiguity  in  the definition of ${\bf \Phi}(z)$
in Eq.~(\ref{eq:defPhi}). In fact, ${\bf \Phi}'(z) = {\bf
\Phi}(z)\,{\cal V}$ satisfies the  same relation as
Eq.~(\ref{eq:defPhi}), so that ${\bf \Phi}(z)$ and ${\bf
\Phi}(z)\,{\cal V}$ must be identified. Since the two matrices ${\bf
Z}$ and ${\bf \Psi}$ are obtained from  $z\,  {\bf \Phi}(z) = {\bf
\Phi}(z) \, {\bf Z} + P(z)\, {\Psi}$, we reach the ${\cal
V}$-equivalence relation,  Eq.~(\ref{eq:VZPsi}).

As an example, let us reproduce the transition function from the
${\cal U}^{(1,1)}$ patch to the ${\cal U}^{(0,2)}$ patch,
Eq.~(\ref{eq:02to11}). The transition matrix from $({\bf
\Psi}^{(1,1)},{\bf Z}^{(1,1)})$ to $({\bf \Psi}^{(0,2)},{\bf
Z}^{(0,2)})$ is ${\cal V}^{-1} = \left(\begin{smallmatrix}b &
a\\1&0\end{smallmatrix}\right)$. One can easily show that the
transition function  Eq.~(\ref{eq:02to11}) is the equivalent to
${\bf Z}^{(0,2)} = {\cal V} \, {\bf Z}^{(1,1)} \, {\cal V}^{-1}$:
\beq
\left(
\begin{array}{cc}
0 & 1 \\
\beta & \alpha
\end{array}
\right)
=
\left(
\begin{array}{cc}
\tilde\phi + b \,\tilde \eta & a \,\tilde\eta\\
\frac{\eta + b\,\phi - b\left(\tilde\phi + b\,\tilde \eta\right)}{a}
& \phi - b\,\tilde\eta
\end{array}
\right).
\eeq

${\bf Z}$ and ${\bf \Psi}$ have a simple physical meaning. First note that the relation
\beq
\det \left(z {\bf 1} - {\bf Z}\right) = P(z)
\eeq
with $P(z) = \det H_0(z)$. This means that zeros $z_i$ of $P(z)$,
namely the position of the vortices, can be obtained as eigenvalues
of the matrix ${\bf Z}$. As an example, let us diagonalize the
matrix ${\bf Z}^{(0,2)} \to {\cal V} \, {\bf Z}^{(0,2)}\, {\cal
V}^{-1} = \left( \begin{smallmatrix} z_1 & 0 \\ 0 & z_2
\end{smallmatrix} \right) $ by ${\cal V}^{-1} =
\left(\begin{smallmatrix} 1 & 1 \\ z_1 & z_2
\end{smallmatrix}\right)$ with $\alpha = z_1 + z_2,\ \beta =
-z_1 \,z_2$. At the same time the other matrix ${\bf \Psi}^{(0,2)}$
is transformed as follows
\beq
{\bf \Psi}^{(0,2)} \to {\bf \Psi}^{(0,2)} \, {\cal V}^{-1} = \left(
\begin{array}{cc}
a \, z_1 + b & a \, z_2 + b\\
1 & 1
\end{array}
\right).
\eeq
The column vectors $\vec\phi_1 = \left(\begin{smallmatrix}a \, z_1 +
b
\\ 1 \end{smallmatrix}\right)$ and $\vec\phi_2 =
\left(\begin{smallmatrix}a \, z_2 + b \\ 1
\end{smallmatrix}\right)$ are nothing but the orientational vectors
given in Eq.~(\ref{eq:vtx:fourkinds}) which are defined at the
vortex positions. We conclude that the eigenvalues of the matrix
${\bf Z}$ are the positions (of the center) of the vortices while
${\bf \Psi}$ is related to the ``orientation'' of the vortices
defined there.

When the two vortex centers coincide, the matrices reduce to the following form:
\beq
\left(
\begin{array}{c}
{\bf \Psi}\\
{\bf Z}
\end{array}
\right)
=
\left(
\begin{array}{cc}
b & a\\
1 & 0\\
0 & 1\\
0 & 0
\end{array}
\right),\
\left(
\begin{array}{cc}
1 & 0\\
0 & 1\\
-X \, Y & X^2\\
-Y^2 & X \,Y
\end{array}
\right),\
\left(
\begin{array}{cc}
1 & 0\\
b' & a'\\
0 & 1\\
0 & 0
\end{array}
\right),
\label{eq:HofN2k2-3}
\eeqq

\subsection{The case of $U(3)$}
Let us next check  the correspondence between the result obtained
above in terms of the matrices ${\bf Z}$ and ${\bf \Psi}$ and the
result from the viewpoint of the moduli matrix $H_0(z)$, for $U(3)$.
There are six patches $\tilde{\cal U}^{(2,0,0)}$, $\tilde{\cal
U}^{(0,2,0)}$, $\tilde{\cal U}^{(0,0,2)}$, $\tilde{\cal
U}^{(1,1,0)}$, $\tilde{\cal U}^{(0,1,1)}$ and $\tilde{\cal
U}^{(1,0,1)}$ for $k=2$ co-axial vortices in the $U(3)$ gauge
theory. These are given by
\begin{alignat}{3}
&H_0^{(0,0,2)}(z) =
\left(
\begin{array}{ccc}
1 & 0 & - a_1 \, z - b_1 \\
0 & 1 & - a_2 \, z - b_2 \\
0 & 0 & z^2
\end{array}
\right),
&\quad
&H_0^{(1,1,0)}(z) =
\left(
\begin{array}{ccc}
z + X\, Y & - {X}^2 & 0 \\
{Y}^2 & z - X\, Y & 0 \\
-\gamma & -\chi & 1
\end{array}
\right),\\
&H_0^{(0,2,0)}(z) =
\left(
\begin{array}{ccc}
1 & - a'_1 \, z - b'_1 & 0 \\
0 & z^2 & 0 \\
0 & - a'_2 \, z - b'_2 & 1
\end{array}
\right),
&\quad
&H_0^{(1,0,1)}(z) =
\left(
\begin{array}{ccc}
z + X'\,Y' & 0 & - {X'}^2 \\
- \gamma' & 1 &  -\chi'\\
{Y'}^2 & 0 & z - X'\,Y'
\end{array}
\right),\\
&H_0^{(2,0,0)}(z) =
\left(
\begin{array}{ccc}
z^2 & 0 & 0 \\
- a''_1 \, z - b''_1 & 1 & 0  \\
- a''_2 \,z - b''_2 & 0 & 1
\end{array}
\right),
&\quad
&H_0^{(0,1,1)}(z) =
\left(
\begin{array}{ccc}
1 & - \gamma'' & - \chi'' \\
0 & z + X''\, Y'' & - {X''}^2 \\
0 & {Y''}^2 & z - X''\, Y''
\end{array}
\right),
\end{alignat}
with identifications
$\left(X,Y\right)\sim \left(-X,-Y\right)$,
$\left(X',Y'\right)\sim \left(-X',-Y'\right)$ and
$\left(X'',Y''\right)\sim \left(-X'',-Y''\right)$.
The determinant of these matrices is equal to $z^2$,  meaning  that
two vortices are sitting on the origin of $z$ plane.

The corresponding matrices ${\bf Z}$ which is $2\,(=k)$ by $2\,(=k)$
matrix and ${\bf \Psi}$ which is $3\,(=N)$ by $2\,(=k)$ matrix for
these moduli matrices can be constructed through the relation
\beq
z \, {\bf \Phi}(z) = {\bf \Phi}(z) \, {\bf Z} + P(z) \, {\bf \Psi}
\label{eq:ZPsi_general}
\eeq
with $P(z) = z^2$.
Here ${\bf \Phi}(z)$ is the orientational matrix defined by
\beq
H_0(z) \, {\bf \Phi}(z) \equiv 0 \label{eq:H0P=0}
\eeq
with modulo $P(z)=z^2$.
We start with the patch $H^{(1,1,0)}(z)$.
The orientational matrix ${\bf \Phi}$ can be chosen as follows
\beq
{\bf \Phi}^{(1,1,0)}(z) =
\left(
\begin{array}{ccc}
z - X\,Y & X^2 \\
- Y^2 & z + X\,Y\\
\gamma z - Y(\gamma \, X + \chi \, Y) & \chi \, z - X(\gamma \, X +
\chi \, Y)
\end{array}
\right).
\eeq
It can be  verified  that this ${\bf \Phi}^{(1,1,0)}(z)$ actually satisfies
the equation Eq.~(\ref{eq:H0P=0}).
According to the equation Eq.~(\ref{eq:ZPsi_general}),
 the matrices ${\bf Z}$ and ${\Psi}$ can be found
as follows:
\beq
z \, {\bf \Phi}^{(1,1,0)}(z) =
{\bf \Phi}^{(1,1,0)}(z)
\left(
\begin{array}{cc}
- X\, Y & X^2\\
-Y^2 & X\, Y
\end{array}
\right)
+ z^2
\left(
\begin{array}{cc}
1 & 0 \\
0 & 1 \\
\gamma & \chi
\end{array}
\right)
\eeq
so \beq
{\bf Z}^{(1,1,0)} =
\left(
\begin{array}{cc}
- X \, Y & X^2\\
-Y^2 & X\, Y
\end{array}
\right)
=
\epsilon
\left(
\begin{array}{c}
-Y\\
X
\end{array}
\right)
\left(
\begin{array}{cc}
-Y & X
\end{array}
\right)
,\quad
{\bf \Psi}^{(1,1,0)} =
\left(
\begin{array}{cc}
1 & 0 \\
0 & 1 \\
\gamma & \chi
\end{array}
\right),
\eeq
with
$\epsilon =
\left(
\begin{smallmatrix}
0 & 1 \\
-1 & 0
\end{smallmatrix}
\right)
$. Then the 2 by 4 matrix $M$ given in Eq.~(\ref{eq:defM}) is of the form
\beq
M^{(1,1,0)}
= \left(
\begin{array}{cccc}
1 & 0 & \gamma & -Y \\
0 & 1 & \chi & X
\end{array}
\right).
\eeq
Similarly  other matrices
corresponding to $H_0^{(1,0,1)}(z)$ and $H_0^{(0,1,1)}(z)$ can be found:
\beq
M^{(1,0,1)} =
\left(
\begin{array}{cccc}
1 & \gamma' & 0 & -Y'\\
0 & \chi' & 1 & X'
\end{array}
\right),\quad
M^{(0,1,1)} =
\left(
\begin{array}{cccc}
\gamma'' & 1 & 0 & -Y''\\
\chi'' & 0 & 1 & X''
\end{array}
\right).
\eeq
Let us next move to the other patch $H^{(0,0,2)}(z)$.
The corresponding orientational matrix ${\bf \Phi}$ is given by
\beq
{\bf \Phi}^{(0,0,2)}(z) =
\left(
\begin{array}{ccc}
b_1\, z & a_1\, z + b_1 \\
b_2\, z & a_2\, z + b_2\\
z & 1
\end{array}
\right).
\eeq
One can verify that this ${\bf \Phi}^{(0,0,2)}(z)$ actually satisfies
the equation Eq.~(\ref{eq:H0P=0}).
According to the equation Eq.~(\ref{eq:ZPsi_general}),
we find the matrices ${\bf Z}$ and ${\Psi}$
as follows
\beq
z {\bf \Phi}^{(0,0,2)}(z) =
{\bf \Phi}^{(0,0,2)}(z)
\left(
\begin{array}{cc}
0 & 1\\
0 & 0
\end{array}
\right)
+ z^2
\left(
\begin{array}{ccc}
b_1 & a_1 \\
b_2 & a_2\\
1 & 0
\end{array}
\right),
\eeq
\beq
{\bf Z}^{(0,0,2)} =
\left(
\begin{array}{cc}
0 & 1\\
0 & 0
\end{array}
\right),\
{\bf \Psi}^{(0,0,2)} =
\left(
\begin{array}{ccc}
b_1 & a_1 \\
b_2 & a_2\\
1 & 0
\end{array}
\right)
\quad \Rightarrow \quad
M^{(0,0,2)}
= \left(
\begin{array}{cccc}
b_1 & b_2 & 1 & 0 \\
a_1 & a_2 & 0 & 1
\end{array}
\right).
\eeq
Similarly,
\beq
M^{(0,2,0)}
= \left(
\begin{array}{cccc}
b_1' & 1 & b_2' & 0 \\
a_1' & 0 & a_2' & 1
\end{array}
\right),\quad
M^{(2,0,0)}
= \left(
\begin{array}{cccc}
1 & b_1'' & b_2'' & 0 \\
0 & a_1'' & a_2'' & 1
\end{array}
\right).
\eeq
These are summarized as follows
\beq
\left(
\begin{array}{c}
\tau_{12}\\
\tau_{23}\\
\tau_{13}\\
\tau_{14}\\
\tau_{24}\\
\tau_{34}
\end{array}
\right)
&\sim&
\left(
\begin{array}{c}
1\\
-\gamma\\
\chi\\
X\\
Y\\
\gamma \, X + \chi \, Y
\end{array}
\right)
\sim
\left(
\begin{array}{c}
\chi'\\
\gamma'\\
1\\
X'\\
\gamma' \, X' + \chi' \, Y'\\
Y'
\end{array}
\right)
\sim
\left(
\begin{array}{c}
-\chi''\\
1\\
\gamma''\\
\gamma'' \, X'' + \chi'' \, Y''\\
X''\\
Y''
\end{array}
\right)
\nonumber\\
&\sim&
\left(
\begin{array}{c}
b_1 \, a_2 - b_2 \, a_1\\
-a_2\\
-a_1\\
b_1\\
b_2\\
1
\end{array}
\right)
\sim
\left(
\begin{array}{c}
-a_1'\\
a_2'\\
b_1'\, a_2' - b_2'\, a_1'\\
b_1'\\
1\\
b_2'
\end{array}
\right)
\sim
\left(
\begin{array}{c}
a_1''\\
b_1''\, a_2'' - b_2''\, a_1''\\
a_2''\\
1\\
b_1''\\
b_2''
\end{array}
\right).
\eeq
The transition functions between these can be easily found via
the weighted equivalence relation Eq.~(\ref{eq:equiv42}).
For example, the transition function from $M^{(1,1,0)}$
to $M^{(1,0,1)}$ is given by $\lambda = \chi^{-\frac{1}{2}}$:
\beq
\chi' = \frac{1}{\chi},\quad \gamma' = - \frac{\gamma}{\chi},\quad
X' = \frac{X}{\chi^{\frac{1}{2}}},\quad Y' = \frac{\gamma \,X +
\chi\, Y}{\chi^{\frac{1}{2}}}.
\eeq
Similarly, the transition function from $M^{(1,1,0)}$
to $M^{(0,2,0)}$ is given by $\lambda = Y^{-1}$:
\beq
a_1' = - \frac{1}{Y^2},\quad a_2' = - \frac{\gamma}{Y^2},\quad b_1'
= \frac{X}{Y},\quad b_2' = \frac{\gamma \, X + \chi \, Y}{Y}.
\eeq
All  other transition functions can be obtained in an analogous way.

\subsection{The case of $U(N)$}
In the general $U(N)$ case there are always two kind of patches for
the moduli matrix $H_0(z)$: $N$ patches with a $z^2$ factor on the
diagonal, which we denote $\tilde{\cal U}^{(i)}$ ($i$ indicates the
position of $z^2$ on the diagonal) and $N(N-1)/2$ patches with 2
diagonal elements of the form $z-c$, which we denote $\tilde{\cal
U}^{(j,k)}$, $j < k$ ($j,k$ indicate the position of the nontrivial
diagonal elements). More explicitly \beq & H_0^{(i)}(z) = \left(
\begin{array}{ccccccc}
1 & 0 & \cdots & 0 & - a_1^{(i)} \, z - b_1^{(i)} & 0 & \cdots \\
0 & 1 & & & \vdots & &\\
\vdots & & \ddots &  & - a_{i-1}^{(i)} \, z - b_{i-1}^{(i)} & & \\
\vdots & & & &  z^2 & & \\
\vdots &&&& - a_{i+1}^{(i)} \, z - b_{i+1}^{(i)} & & \\
\vdots & & & & \vdots & &
\end{array}
\right)  \label{genform1} \eeq \beq & H_0^{(j,k)}(z) = \left(
\begin{array}{ccccccccccc}
1 & 0 & \cdots & 0 & -\gamma_1^{(jk)} & 0 & \cdots & 0 &
-\chi_1^{(jk)} & 0 & \cdots \\
0 & 1 & & & \vdots & & & & \vdots & & \\
\vdots &  & \ddots &  & -\gamma_{j-1}^{(jk)} &  &  &  &
-\chi_{j-1}^{(jk)} &  &  \\
\vdots & & & & z + X^{(jk)} \, Y^{(jk)} & & & & - (X^{(jk)})^2 &  & \\
\vdots &  &  &  & -\gamma_{j+1}^{(jk)} &  &  &  &
-\chi_{j+1}^{(jk)} &  &  \\
\vdots & & & & \vdots & & & &\vdots & & \\
\vdots &  &  &  & -\gamma_{k-1}^{(jk)} & & & &
-\chi_{k-1}^{(jk)} &  &  \\
\vdots & &&& (Y^{(jk)})^2 &&&& z - X^{(jk)} \, Y^{(jk)} & &\\
\vdots &  &  &  & -\gamma_{k+1}^{(jk)} &  &  &  &
-\chi_{k+1}^{(jk)} &  &  \\
\vdots & & & & \vdots & & & &\vdots & &
\end{array}
\right) \label{genform2}
\eeqq
It is possible to find the transition functions among the moduli in
the different patches using appropriate $V$-transformations. It
turns out that only a subset of transition functions is needed, then
the others can be recovered using composition and inversion of the
known ones. In particular we need
\begin{itemize}
\item $\tilde{\cal U}^{(i)}\to \tilde{\cal U}^{(j)}$
\beq && a_k^{(j)} =  {a_k^{(i)} \, b_j^{(i)} - a_j^{(i)} \,
b_k^{(i)} \over (b_j^{(i)})^2}, \quad k\ne i;  \qquad
 a_i^{(j)}=-{a_j^{(i)} \over (b_j^{(i)})^2} \\
 && b_k^{(j)}= {b_k^{(i)} \over b_j^{(i)}}, \quad k\ne i; \qquad
b_i^{(j)}= {1\over b_j^{(i)}}
\eeqq

\item $\tilde{\cal U}^{(i)}\to \tilde{\cal U}^{(k,i)}$
\beq && X^{(ki)}=\pm i {b_k^{(i)} \over \sqrt{a_k^{(i)}}};  \qquad
Y^{(ki)} =
\pm {1\over\sqrt{a_k^{(i)}}} \\
&& \gamma_l^{(ki)}= {a_l^{(i)} \over a_k^{(i)}};  \qquad
\chi_l^{(ki)}= b_l^{(i)}-{a_l^{(i)} \, b_k^{(i)} \over a_k^{(i)}}
\eeqq

\item $\tilde{\cal U}^{(i)}\to \tilde{\cal U}^{(j,k)}$, $j,k\ne i$
\beq && X^{(jk)}=\pm {b_j^{(i)} \over \sqrt{d_{jk}^{(i)}}};  \qquad
Y^{(jk)} = \pm {b_k^{(i)} \over \sqrt{d_{jk}^{(i)}}}\\
&& \gamma_l^{(jk)}= {a_k^{(i)} \, b_l^{(i)}-a_l^{(i)} \, b_k^{(i)} \over  \sqrt{-d_{jk}^{(i)}}}, \quad l\ne i;  \qquad \gamma_i^{(jk)}= {a_k^{(i)} \over  \sqrt{-d_{jk}^{(i)}}} \\
&& \chi_l^{(jk)}= {a_j^{(i)} \, b_l^{(i)}-a_l^{(i)} \, b_j^{(i)}
\over \sqrt{d_{jk}^{(i)}}}, \quad l\ne i;  \qquad \chi_i^{(jk)}=
{a_j^{(i)}
\over  \sqrt{d_{jk}^{(i)}}} \\
\eeqq
with $d_{jk}^{(i)}\equiv -a_j^{(i)} \, b_k^{(i)}+a_k^{(i)} \,
b_j^{(i)}$.

\end{itemize}

On the other hand, given the general form of $H_0(z)$,
Eq.~(\ref{genform1}) and  Eq.~(\ref{genform2}), the moduli can
always be collected into the 2 by $N+1$ matrix $M=({\bf \Psi}^T, v)$
with the usual procedure (${\bf Z}= \epsilon \, v \, v^T$). If we
denote $M^{(i)}$, $M^{(jk)}$ in the patches $\tilde{\cal U}^{(i)},
\tilde{\cal U}^{(j,k)}$ respectively we get \beq & M^{(i)}= \left(
\begin{matrix}
b_1^{(i)} &  \cdots & b_{i-1}^{(i)} &1 & b_{i+1}^{(i)} & \cdots & b_N^{(i)} & 0\\
a_1^{(i)} &\cdots & a_{i-1}^{(i)} & 0 & a_{i+1}^{(i)} & \cdots &
a_N^{(i)} & 1
\end{matrix} \right) \\
& M^{(jk)} = \left(
\begin{array}{cccccccccccc}
\gamma_1^{(jk)} & \cdots & \gamma_{j-1}^{(jk)} & 1 &
\gamma_{j+1}^{(jk)} & \cdots & \gamma_{k-1}^{(jk)} & 0 &
\gamma_{k+1}^{(jk)} & \cdots & \gamma_N^{(jk)} & -Y^{(jk)} \\
\chi_1^{(jk)} & \cdots & \chi_{j-1}^{(jk)} & 0 & \chi_{j+1}^{(jk)} &
\cdots & \chi_{k-1}^{(jk)} & 1 & \chi_{k+1}^{(jk)} & \cdots &
\chi_N^{(jk)} & X^{(jk)}
\end{array} \right)
\eeqq The matrix $M$ together with the weighted $GL(2,{\bf C})$,
Eq.~(\ref{glquot}), defines the weighted Grassmannian manifold
$WGr_{N+1,2}^{(1,\ldots,1,0)}$ and the $M^{(i)}$, $M^{(jk)}$
represent the standard covering of this space. One can pass from one
patch to another by appropriate weighted $GL(2,{\bf C})$
transformation and so deduce the transition functions, which turn
out to be the same of the moduli matrix representation, as expected.
In particular it is possible to check that the transition functions
listed above, which generate all the others, perfectly match with
the corresponding ones of the $WGr_{N+1,2}^{(1,\ldots,1,0)}$.

We have thus explicitly pointed out that the moduli space of $k=2$
vortices given by the moduli matrix $H_0(z)$ is indeed a weighted
Grassmannian manifold $WGr_{N+1,2}^{(1,\ldots,1,0)}$. This enforces
the general considerations coming from the established  equivalence
between the moduli matrix and the K\"ahler quotient construction
\cite{Eto:2006pg}.

\section{Product of moduli matrices}

Within the moduli matrix formalism, it is easy to construct vortices
of higher winding number: the latter can be constructed from the
moduli matrices of lower winding number  as simple products. For
instance, consider two fundamental  vortices, and
\beq H_{0}^{(1,0)}
\times  H_{0}^{(1,0) \, \prime} = \left(\begin{array}{cc}z-z_0 & 0
\\
 -b_{0}  & 1\end{array}\right)\,
\left(\begin{array}{cc}z-z_0^{\prime}  & 0 \\  -b_{0}^{\prime}  &
1\end{array}\right) =  \left(\begin{array}{cc}(z-z_0)
(z-z_0^{\prime}) & 0 \\
-b_0 \, z + b_0 \, z_0^{\prime} - b_0^\prime & 1\end{array}\right).
\eeqq
Analogously for the product of two $(0,1)$ vortices
\beq
H_{0}^{(0,1)} \times  H_{0}^{(0,1) \, \prime} =
\left(\begin{array}{cc}1 & -a_{0} \\ 0  & z-z_0
\end{array}\right)\, \left(\begin{array}{cc}  1 &   -a_{0}^{\prime}
\\ 0 & z-z_0^{\prime}\end{array}\right) =  \left(\begin{array}{cc}1
& -a_0 \, z + a_0 \, z_0^{\prime} - a_0^{\prime}  \\ 0 & (z-z_0)
(z-z_0^{\prime}) \end{array}\right).
\eeqq
By comparing these  with
$ H_{0}^{(0,2)} $ or  $ H_{0}^{(2,0)} $  in  Eq.~(\ref{eq:HofN2k2}),
one finds
\beq a= a_{0}, \quad  b= a_0 \,  z_0^{\prime} - a_0^{\prime},
\quad \alpha= z_0 + z_0^{\prime}, \quad  \beta= -   z_0  \,
z_0^{\prime}; \label{compos1}
\eeqq
\beq a^{\prime}= -b_{0}, \quad
b^{\prime}= b_0 \, z_0^{\prime} - b_0^{\prime}, \quad \alpha= z_0 +
z_0^{\prime}, \quad  \beta= -   z_0  \, z_0^{\prime}.
\label{compos2}
\eeqq
Finally, for the product vortex of the type $(0,1)$ times $(1,0),$
\beq   H_{0}^{(1,0)} \times H_{0}^{(0,1) \, \prime} =
\left(\begin{array}{cc}z-z_0 & 0 \\  -b_{0}  & 1\end{array}\right)\,
\left(\begin{array}{cc}  1 &   -a_{0}^{\prime}
\\ 0 & z-z_0^{\prime}\end{array}\right) =
\left(\begin{array}{cc}z-z_0 & -a_0^{\prime} \, (z-z_0)    \\-b_0 &
z - z_0^{\prime}  + b_0 \, a_0^{\prime}\end{array}\right).
\eeqq
Bringing it to the standard form by a $V(z)$ transformation
\beq  V=
\left(\begin{array}{cc}1 & a_0^{\prime} \\0 & 1\end{array}\right),
\eeqq
one has
\beq H_{0}^{(1,1)} \sim    H_{0}^{(1,0)} \times
H_{0}^{(0,1) \, \prime} \simeq  \left(\begin{array}{cc} z-z_0 - b_0
\, a_0^{\prime} &  a_0^{\prime}\, ( z_{0}- z_{0}^{\prime})  +
a_0^{\prime  2} \, b_{0} \\ -b_0 & z - z_0^{\prime} +b_0 \,
a_0^{\prime}\end{array}\right).
\eeqq
This has the same form as the middle form of Eq.~(\ref{eq:HofN2k2}),
by identification
\beq \eta=   - a_0^{\prime}\, ( z_{0}-
z_{0}^{\prime})   -  a_0^{\prime 2} \, b_{0}, \quad  {\tilde \eta} =
b_0, \quad \phi =  z_0 + b_0 \, a_0^{\prime}, \quad  {\tilde \phi} =
z_0^{\prime} -  b_0 \, a_0^{\prime}. \label{identify}
\eeqq
Note that
\beq  \phi +  {\tilde \phi} =   z_0  + z_0^{\prime} = \alpha, \quad
\eta \, {\tilde \eta} - \phi \,  {\tilde \phi} = - z_{0} \,
z_{0}^{\prime}= \beta.
 \eeqq
in accord with the relations  Eq.~(\ref{eq:02to11}).

In fact, the transition function Eq.~(\ref{eq:02to20}) between the
sets $(a,b, \alpha, \beta)$  and  $(\phi, {\tilde \phi}, \eta,
{\tilde \eta})$   (patches ($(0, 2)$ and $(1, 1)$) is   simply a
consequence  of the  transition function  for  the minimum vortex
\beq  b_{0}= \frac{1}{a_{0}}, \eeqq through the composition rule,
Eq.~(\ref{compos1})  and  Eq.~(\ref{identify}).

Analogously, to find  the relation  between the $(2,0)$ and $(1,1)$
patches, we first write  $H_{0}^{(1,1)}$ as
\beq  H_{0}^{(1,0)}
\times H_{0}^{(0,1) \, \prime} =  \left(\begin{array}{cc}  1 &
-a_{0}  \\ 0 & z-z_0 \end{array}\right)  \,  \left(\begin{array}{cc}
z-z_0^{\prime} & 0 \\  -b_{0}^{\prime}  & 1\end{array}\right)\, =
\left(
\begin{array}{cc}
z-\phi & -\eta\\
-\tilde\eta & z-\tilde\phi
\end{array}
\right).\label{eq:1120}
\eeq
The relation
\beq   a^{\prime}=
\frac{1}{\eta};\quad  b^{\prime}= - \frac{\phi}{\eta};  \quad
\alpha =\phi +{\tilde \phi}; \quad \beta=  {\eta}\, {\tilde
\eta} -\phi \, {\tilde \phi}, \label{eq:11to20}
\eeqq
follows then  easily from the transition rule $ b_{0} = 1/ a_{0}$
between $(0,1)$ and $(1,0)$ patches. Finally the relation
Eq.~(\ref{eq:02to20}) follows by composing Eq.~(\ref{eq:02to11}) and
Eq.~(\ref{eq:11to20}).

In the case of co-axis vortices, one gets, by eliminating the
center-of mass position and the relative position (by setting
$z_{0}= z_{0}^{\prime}=0$),
\beq H_{0}^{(1,1)} \sim
\left(\begin{array}{cc} z-\phi  &  - \eta  \\ - {\tilde \eta}  &   z
+ \phi  \end{array}\right), \qquad  \phi^{2} + \eta \, {\tilde
\eta}=0.
\eeqq
Note that in this construction  ($ H_{0}^{(1,1)} \sim H_{0}^{(0,1)}
\times  H_{0}^{(0,1)}$),  the constraint $\phi^{2} + \eta \, {\tilde
\eta}=0$ is automatically satisfied once we set $z_{0}=
z_{0}^{\prime}=0$ due to the identification Eq.~(\ref{identify}).

These discussions simply show  that the moduli matrices have a
natural property under the product. Thus
  \beq   H_{0}^{(m, n)} \sim H_{0}^{(m_{1}, n_{1})} \times H_{0}^{(m_{2}, n_{2})}, \qquad  m_{1}+m_{2}=m; \quad
  n_{1}+ n_{2} = n.
  \eeqq
The product moduli is simply
  \beq {\cal M}_{1}  \times   {\cal M}_{2},
  \eeqq
as long as no constraints (such as the coincident axes) are imposed.
The transition functions between the ``neighboring'' patches  (say
$(m+1, n)$ and  $(m, n+1)$)  can be always reduced to the simple
relation    between  $H_{0}^{(1,0)}$ and $H_{0}^{(0,1)}$.

\newcommand{\J}[4]{{\sl #1} {\bf #2} (#3) #4}
\newcommand{\andJ}[3]{{\bf #1} (#2) #3}
\newcommand{\AP}{Ann.\ Phys.\ (N.Y.)}
\newcommand{\MPL}{Mod.\ Phys.\ Lett.}
\newcommand{\NP}{Nucl.\ Phys.}
\newcommand{\PL}{Phys.\ Lett.}
\newcommand{\PR}{ Phys.\ Rev.}
\newcommand{\PRL}{Phys.\ Rev.\ Lett.}
\newcommand{\PTP}{Prog.\ Theor.\ Phys.}
\newcommand{\hep}[1]{{\tt hep-th/{#1}}}

\end{document}